\documentclass[
onecolumn,superscriptaddress,preprintnumbers,nofootinbib,amsmath,noshowpacs,aps,pra,floatfix,nolongbibliography]{revtex4-2}
\usepackage[utf8]{inputenc}
\usepackage{graphicx}
\usepackage{dcolumn}
\usepackage{bm}
\usepackage{float}
\usepackage{tikz}
\usetikzlibrary{quantikz2}
\usepackage{amssymb}
\usepackage{mathtools}
\usepackage{array}
\usepackage{fixmath}
\usepackage{subcaption}
\usepackage{mathrsfs}
\usepackage{xcolor,colortbl}
\usepackage{bbm}
\usepackage{physics}
\usepackage{multirow}
\usepackage{upgreek}
\usepackage{appendix}
\usepackage{verbatim}
\usepackage[pdftex,colorlinks,linkcolor=black,citecolor=magenta,urlcolor=black]{hyperref}
\usepackage{tabularx}
\usepackage{adjustbox}
\usepackage{rotating}
\usepackage{ragged2e}
\usepackage{algorithm}
\usepackage{algpseudocode}
\usepackage{enumitem}
\usepackage{wrapfig}%wrap figure

\newlist{steps}{enumerate}{1}
\setlist[steps, 1]{label = Step \arabic*:}

\newtheorem{theorem}{Theorem}

\newcommand{\EC}{\textsc{ECPointAdd }}

\newcommand{\firstloopbox}[2]{%
    \gategroup[#1,steps=#2,style={dashed,rounded corners,fill=gray!0,inner xsep=2pt},
    background,label style={label position=above,anchor=north,yshift=0.3cm}]{for $i = 1,\ldots, t$}%
}

\newcommand{\secondloopbox}[2]{%
    \gategroup[#1,steps=#2,style={dashed,rounded corners,fill=gray!0,inner xsep=2pt},
    background,label style={label position=above,anchor=north,yshift=0.3cm}]{Correction Step}%
}

\begin{document}

\title{Quantum resource estimates for computing binary elliptic curve discrete logarithms}

\author{Michael Garn}\thanks{michael.garn@stfc.ac.uk}
\affiliation{The Hartree Centre, STFC, Sci-Tech Daresbury, Warrington, WA4 4AD, UK}

\author{Angus Kan}\thanks{akan@psiquantum.com}
\affiliation{PsiQuantum, 700 Hansen Way, Palo Alto, California 94304, USA}
\affiliation{PsiQuantum, Daresbury, WA4 4FS, UK}

\date{\today}

\begin{abstract}
We perform logical and physical resource estimation for computing binary elliptic curve discrete logarithms using Shor's algorithm on fault-tolerant quantum computers. We adopt a windowed approach to design our circuit implementation of the algorithm, which comprises repeated applications of elliptic curve point addition operations and table look-ups. Unlike previous work, the point addition operation is implemented exactly, including all exceptional cases. We provide exact logical gate and qubit counts of our algorithm for cryptographically relevant binary field sizes. Furthermore, we estimate the hardware footprint and runtime of our algorithm executed on surface-code matter-based quantum computers with a baseline architecture, where logical qubits have nearest-neighbor connectivity, and on a surface-code photonic fusion-based quantum computer with an active-volume architecture, which enjoys a logarithmic number of non-local connections between logical qubits. At 10$\%$ threshold and compared to a baseline device with a $1\mu s$ code cycle, our algorithm runs $\gtrsim$ 2-20 times faster, depending on the operating regime of the hardware and over all considered field sizes, on a photonic active-volume device.
\end{abstract}

\maketitle

%\tableofcontents

\section{Introduction}
Shor's algorithm~\cite{Shor,shor2} computes discrete logarithms for finite Abelian groups in polynomial time and thus, can be used to break RSA cryptography~\cite{rsa}, which is based on the hardness of integer factorization. It can also be applied to elliptic curve groups to efficiently break elliptic curve cryptography (ECC)~\cite{galbraith2016recent}. A large corpus of work has put the theoretical vulnerability of these cryptosystems against quantum attacks in more concrete terms by analyzing and reducing the quantum-computational resources, e.g., gate and qubit counts, that will be sufficient to break them~\cite{proos2004shors,kaye2004optimized,cheung2008design,maslov2009m2,PhysRevA.86.032324,amento2013efficient,Roetteler2017quantum,10.5555/3179553.3179560,gheorghiu2019benchmarkingquantum,van2019space,10.1007/978-3-030-44223-1_23,Banegas2020Concrete,PhysRevLett.131.040602,Gidney2021how,Dedy2022another,litinski2022activevolume,litinski2023compute,Putranto2023Depth,kim2023new,Taguchi2023Concrete,kim2024toffoli,Taguchi2024On}. To safeguard against sufficiently powerful quantum computers, NIST have recently proposed timelines to transition to post-quantum cryptography~\cite{chen2023recommendations,moody2024transition} over the next decade. 

Putting practicality aside, computing discrete logarithms is scientifically intriguing, as it belongs to a very selective class of problems, for which a quantum algorithm achieves a super-polynomial speed-up over any existing classical algorithms and the computed solution is efficiently verifiable classically. In this work, we focus on solving the discrete logarithm problem for elliptic curves over a binary field. In particular, we are interested in, under reasonable assumptions about the quantum computer architecture, estimating (i) the size of a quantum computer that will be sufficient to solve this problem for cryptographically relevant field sizes and (ii) the runtime of such a computation. 

\textbf{Architectural assumptions:} The cryptographically relevant problem sizes well exceed the capability of current non-error-corrected quantum computers; we expect that they can only be solved on fault-tolerant quantum computers (FTQC) enabled by error correction. We will focus on resource estimates for FTQC based on the surface code~\cite{bravyi1998quantumcodeslatticeboundary,KITAEV20032,PhysRevA.86.032324,PhysRevA.71.022316,fowler2019lowoverheadquantumcomputation,Litinski2019gameofsurfacecodes,bombin2021interleavingmodulararchitecturesfaulttolerant}, which encode each logical qubit in a two-dimensional patch of physical qubits. We consider two types of surface-code architectures: (i) baseline architectures where physical and logical qubits communicate via nearest-neighbor operations on a two-dimensional grid~\cite{PhysRevA.86.032324,fowler2019lowoverheadquantumcomputation,Litinski2019gameofsurfacecodes,bombin2021interleavingmodulararchitecturesfaulttolerant}, and (ii) the active-volume (AV) architecture~\cite{litinski2022activevolume} that utilizes a logarithmic amount of non-local connections between logical qubits; these non-local connections facilitate a higher level of parallelization between logical operations, which in turn, bring about a significant speed-up, compared to a baseline FTQC with a similar physical footprint but only two-dimensional, local connectivity.

\textbf{Previous works:} There have been significant efforts over the past two decades in constructing and optimizing quantum algorithms that solve the binary elliptic curve discrete logarithm problem~\cite{proos2004shors,kaye2004optimized,cheung2008design,maslov2009m2,amento2013efficient,Banegas2020Concrete,Dedy2022another,Putranto2023Depth,kim2023new,Taguchi2023Concrete,kim2024toffoli,Larasati2024Quantum,Taguchi2024On}. Particularly germane to this paper are the works that provide explicit resource estimates in terms of logical gate and qubit counts, and optimize for Toffoli count~\cite{Banegas2020Concrete,Taguchi2023Concrete,kim2024toffoli,Taguchi2024On}. This is so because the product of Toffoli count and logical qubit count largely determines the runtime and footprint for baseline architectures~\cite{Litinski2019gameofsurfacecodes,litinski2023compute,Gidney2021how}. However, to our knowledge, unlike existing works on RSA and ECC over a prime field~\cite{gheorghiu2019benchmarkingquantum,Gidney2021how,PhysRevLett.131.040602,litinski2023compute,litinski2022activevolume}, all the resource estimates from existing binary ECC works remain at an abstract, hardware agnostic level, i.e., logical gate and qubit counts, and stop short of estimating physical, hardware-relevant resources, e.g., physical runtime and number of physical qubits for matter-based FTQC, or number of resource-state generators for photonic FTQC.

\textbf{Our contributions:} We provide both abstract and physical resource estimates for both baseline and AV architectures. We focus on superconducting and atomic hardware, including trapped-ion and neutral-atom platforms, for baseline architectures, and photonic fusion-based quantum computing~\cite{bartolucci2023fusion} for the AV architecture. We stress that in principle, baseline and AV architectures are both hardware-agnostic. The association between baseline architectures and matter-based hardware, and the association between AV architecture and fusion-based photonic platforms~\cite{bartolucci2023fusion,PhysRevLett.133.050605,PhysRevLett.133.050604,chan2024tailoringfusionbasedphotonicquantum,bartolucci2021creationentangledphotonicstates,PhysRevA.110.032402} are motivated by practical reasons: Long-range connections between matter-based qubits come with a variety of challenges, e.g., frequency conversion~\cite{PhysRevX.14.031055}, high-quality cavities~\cite{sinclair2024faulttolerantopticalinterconnectsneutralatom,PhysRevA.89.022317}, low-rate Bell measurements~\cite{PhysRevLett.130.050803,storz2023loophole} and slow shuttling~\cite{PhysRevX.14.041028}, whereas low-loss fiber and high-quality photonic-chip-to-fiber coupling could more directly support long-range connections between photonic qubits hosted on separate chips~\cite{alexander2024manufacturableplatformphotonicquantum, AghaeeRad2025}. Following~\cite{litinski2022activevolume,litinski2023compute}, we estimate the number of physical qubits and runtime on matter-based platforms, and estimate the number of resource-state generators and runtime on a photonic fusion-based quantum computer (FBQC), called for by our algorithm.

Our other contributions include pedagogical reviews of recent advances in binary-field arithmetic quantum circuits~\cite{kim2024toffoli,Taguchi2023Concrete,Taguchi2024On}, which implement known classical algorithms~\cite{itoh1988fast,sunar2004generalized,fan2007comments,find2018better,ccalik2019searching} and are used in our algorithm, as well as optimizations and necessary corrections therein. Furthermore, our binary elliptic curve point addition routine incorporates all exceptional cases of the point addition operation, including, e.g., the point-doubling case~\cite{amento2013efficient,galbraith2016recent}, which has been previously attempted~\cite{Larasati2024Quantum} though not accomplished.

The rest of the paper is organized as follows: In section~\ref{sec:binaryEC}, we provide necessary background on binary ECC. In section~\ref{sec:algo}, we provide an overview of our algorithm and the subroutines employed therein, supplemented by materials in the appendices. In section~\ref{sec:QRE}, we present the methods used to estimate the abstract and physical computational resources, and our estimates for relevant binary-field sizes. We summarize our findings and discuss future directions in section~\ref{sec:discussion}.

\section{Binary Elliptic Curves}
\label{sec:binaryEC}
Binary elliptic curves are elliptic curves defined over a binary field $\mathbbm{F}_{2^n}$. We use a polynomial basis representation: $\mathbbm{F}_{2^n}$ is identified with $\mathbbm{F}_2[x]/p(x)$, where $p(x) \in \mathbbm{F}_2[x]$ is an irreducible polynomial of degree $n$. Then, the elements in $\mathbbm{F}_{2^n}$ are represented as polynomials of degree less than $n$ with binary coefficients in $\mathbbm{F}_2$. All computations are done modulo $p(x)$. We adopt polynomials $p(x)$ that are used in the standardized binary elliptic curves listed in~\cite{kerry2013digital} and displayed in table~\ref{tab:irreducible_polynomials}.

An ordinary binary elliptic curve is given by
\begin{equation}\label{eq:curve}
    y^2 + xy = x^3 + ax^2 + b,
\end{equation}
where $a \in \mathbbm{F}_{2^n}$ and $b \in \mathbbm{F}_{2^n}^*$. The set of points on a curve consist of tuples $P = (x,y) \in \mathbbm{F}^2_{2^n}$, which satisfy equation~\eqref{eq:curve}, and the so-called point at infinity $\mathcal{O}$. This set forms a group under point addition, where given $P_1=(x_1, y_1)$ and $P_2=(x_2, y_2)$, $P_3 = (x_3, y_3)$ is conventionally given by~\cite{amento2013efficient}
\begin{gather}
(x_3, y_3) = 
    \begin{cases} \label{eq:special_case1}
        \mathcal{O} &\text{if $P_1 = -P_2$,} \\
        (x_1, y_1) &\text{if $P_2 = \mathcal{O}$,} \\
        (x_2, y_2) &\text{if $P_1 = \mathcal{O}$,}
    \end{cases} \\
\text{ else if } P_1 = P_2, \:(x_3, y_3) = (\lambda^2 + \lambda + a, x_2^2 + (\lambda + 1) x_3) \: \text{ with } \lambda = x_2 + \frac{y_2}{x_2}, \label{eq:point_doubling} \\
\text{ else if } P_1 \neq \pm P_2, \:(x_3, y_3) = (\lambda^2 + \lambda + x_1 + x_2 + a, (x_2 + x_3)\lambda +x_3 + y_2) \: \text{ with } \lambda = \frac{y_1 + y_2}{x_1 + x_2}.\label{eq:last_case} 
\end{gather}
Here, $-P_2 = (x_2, y_2 + x_2)$. We choose $(0,0)$ as the representation of $\mathcal{O}$~\cite{amento2013efficient}. We work with a recasted form of elliptic curve point addition, which we hereafter abbreviate to \textsc{ECPointAdd}; we rewrite~\eqref{eq:point_doubling} and~\eqref{eq:last_case} in the following form:
\begin{gather}
    \text{ else } (x_3, y_3) = (\lambda^2 + \lambda + x_1 + x_2 + a, (x_2 + x_3)\lambda +x_3 + y_2), \nonumber \\
    \text{ with } \lambda = \begin{cases}
        \lambda_r = x_2 + \frac{y_2}{x_2} =\frac{y_3 + x_3 + y_2}{x_2 + x_3} &\text{ if $P_1 = P_2$,} \\
        \frac{y_1 + y_2}{x_1 + x_2}  =\frac{y_3 + x_3 + y_2}{x_2 + x_3} &\text{ otherwise.}
    \end{cases}\label{eq:combined}
\end{gather}
Using the fact that each polynomial is represented as a bit-string, where the $i$th bit is the $i$th polynomial coefficient, and that adding two polynomials is done via bitwise XOR, as we explain in section~\ref{sec:arith} and appendix~\ref{appendix:toff_free}, one can show that~\eqref{eq:combined} is equivalent to~\eqref{eq:point_doubling} and~\eqref{eq:last_case}. Consider first the point-doubling case $P_1 = P_2$, i.e.,~\eqref{eq:point_doubling}. $x_3 \stackrel{(3)}{=} \lambda^2 + \lambda + a = \lambda^2 + \lambda + x_1 + x_2 + a$, as claimed in~\eqref{eq:combined}, because $x_1 + x_2 = 0$. Next, $y_3 \stackrel{(3)}{=} x_2^2 + (\lambda + 1) x_3 = \lambda x_2 + y_2 + (\lambda + 1) x_3 = (x_2 + x_3)\lambda +x_3 + y_2$, as claimed in~\eqref{eq:combined}, because $x_2^2 = (x_2 + \frac{y_2}{x_2}) x_2 +y_2 \stackrel{(3)}{=} \lambda x_2 + y_2$. Finally, $y_3 = (x_2 + x_3)\lambda +x_3 + y_2$ implies that $\lambda = \frac{y_3 + x_3 + y_2}{x_2 + x_3}$, as claimed in~\eqref{eq:combined}. The case where $P_1 \neq \pm P_2$, i.e.,~\eqref{eq:last_case}, is now manifestly handled.

The Diffie-Hellman key-exchange mechanism and security of binary ECC rely on the fact that while a sum of $k$ $P$'s under point addition, denoted hereafter by $Q = [k]P$, can be computed classically in polynomial time via~\eqref{eq:point_doubling}, there is no known polynomial-time classical algorithm that computes $k$ (private key) from $P$ (base point) and $Q$ (public key). This problem is known as the binary elliptic curve discrete logarithm problem (ECDLP). For more background on ECC, consult, e.g.,~\cite{cohen2005handbook}.

\section{Algorithm and Subroutines}
\label{sec:algo}

In this section, we present our construction of Shor's algorithm for the binary ECDLP. We start by reviewing the high-level structure of the algorithm, and then proceed to break it down into fundamental subroutines, among which, the binary-field arithmetic routines are discussed in detail.

\subsection{Algorithm structure}
\label{sec:alg_struct}

\begin{figure}[!ht]
    \centering
    \includegraphics[width=\linewidth]{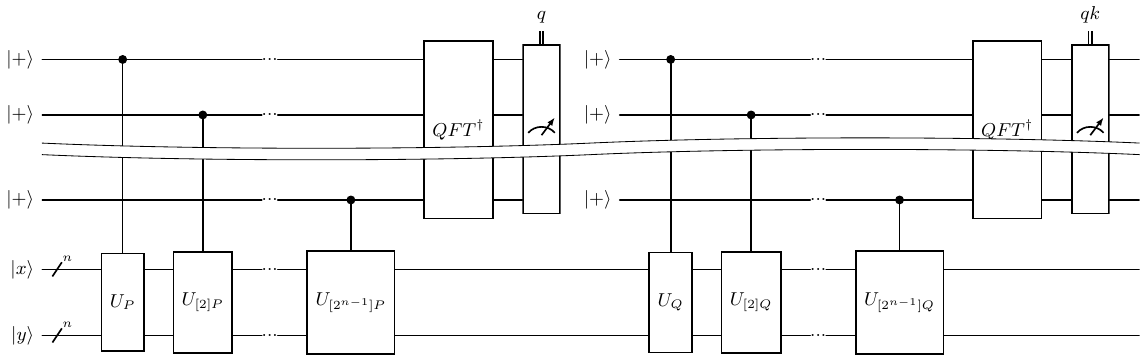}
    \caption{Shor's algorithm circuit that computes the private key $k$ from the base point $P=(x,y)$ and public key $Q=[k]P$.}
    \label{fig:Shor}
\end{figure}

\begin{figure}[ht!]
    \centering
    \begin{adjustbox}{width=\textwidth}
    \begin{quantikz}
            \lstick{$\ket{+}^{\otimes s}$} & \qwbundle{k} & \ctrl{1} & \ctrl{1}& \push{...}&\ctrl{1} & \\
            \lstick{$\ket{x_1}$} & \qwbundle{n} & \gate[2]{U_R} &\gate[2]{U_{[2]R}}& \push{...}& \gate[2]{U_{[2^{s}]R}}&\\
            \lstick{$\ket{y_1}$} & \qwbundle{n} &  & & \push{...}& &
    \end{quantikz}
    =
    \begin{quantikz}
        \lstick{$\substack{\ket{+}^{\otimes s}  \propto \\ \sum_{q=0}^{2^s-1} \ket{q}}$} & \qwbundle{k} & \gate[1]{\mbox{Input }q}\wire[d][3]{q}&&\gate[1]{\mbox{Input }q}\wire[d][3]{q}&\\
        \lstick{$\ket{x_1}$} & \qwbundle{n} & &\gate[5]{\EC}&&\\
        \lstick{$\ket{y_1}$} & \qwbundle{n} & &&&\\
        \lstick{$\ket{x_2=0}$} & \qwbundle{n} &\gate[2]{\mbox{Look-up }(x_2,y_2) = [q]R} &&\gate[2]{\mbox{Un-look-up}}&\\
        \lstick{$\ket{y_2=0}$} & \qwbundle{n} &\wire[d][1]{q}&&\wire[d][1]{q}&\\
        \lstick{$\ket{0}$} & \qwbundle{n} &\gate[1]{\mbox{Look-up }\lambda_r(x_2,y_2)}&&\gate[1]{\mbox{Un-look-up}}&\\
    \end{quantikz}
    \end{adjustbox}
    \caption{Following~\cite{10.1007/978-3-030-44223-1_23,litinski2023compute}, we implement a group of $s$ controlled-addition of a point $R$ using a look-up table, uncontrolled point addition operation, and an uncomputation of the look-up table. The optimal window size $s$ is calculated for every considered elliptic curve field size.}
    \label{fig:window}
\end{figure}
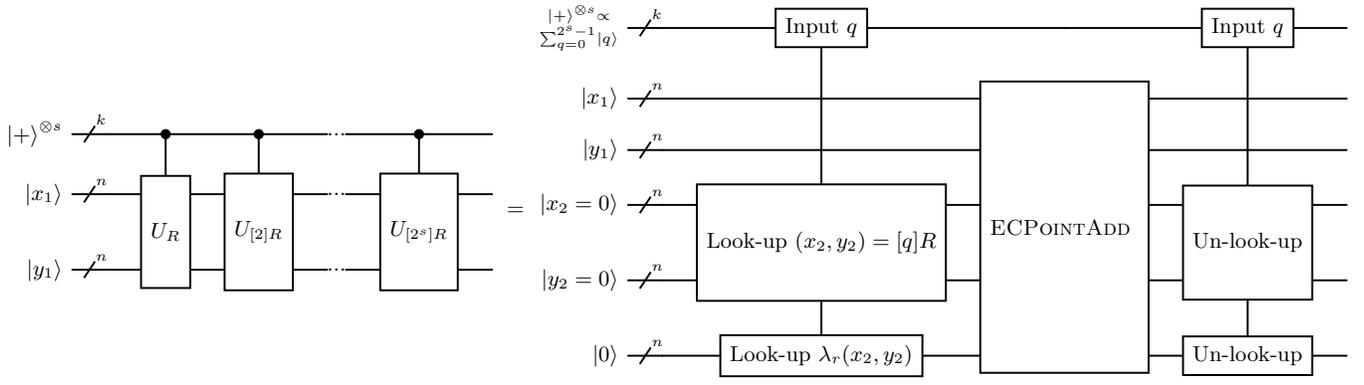

The high-level circuit of Shor's algorithm for computing binary ECDLP is displayed in figure~\ref{fig:Shor}. It consists of two rounds of quantum phase estimation (QPE) applied to the unitaries $U_P$ and $U_Q$, which add the base point $P$ and public key $Q = [k]P$ respectively to any input $(x,y)$. The first QPE is performed on the input state $\ket{P}=\ket{x}\otimes \ket{y}$; it prepares an eigenstate of $U_P$ with non-trivial overlap with $\ket{P}$:
\begin{equation}
    \ket{\phi_q} = \sum_{j=0}^{r-1}e^{iqj\frac{2\pi}{r}}\ket{[j]P},
\end{equation}
whose eigenphase is given by
\begin{equation}
    U_P\ket{\phi_q} = e^{-iq\frac{2\pi}{r}}\ket{\phi_q},
\end{equation}
and outputs a non-negative integer $q < r$, where $r$ is the order of $P$, i.e., $[r]P = P$. $\ket{\phi_q}$ is also an eigenstate of $U_Q$ with
\begin{equation}
    U_Q \ket{\phi_q} = \sum_{j=0}^{r-1}e^{iqj\frac{2\pi}{r}}\ket{[j]P+Q} = \sum_{j=0}^{r-1}e^{iqj\frac{2\pi}{r}}\ket{[j+k]P} = e^{-iqk\frac{2\pi}{r}}\ket{\phi_q}.
\end{equation}
Hence, the second QPE will output $qk$, from which $k$ can be obtained after dividing $qk$ by $q$.

We follow the windowing method from~\cite{10.1007/978-3-030-44223-1_23,Gidney2021how,litinski2023compute} to implement groups of controlled point addition operations. In particular, we first divide the $n$ controlled point additions into groups of $s$ contiguous operations (the last group will contain less than $s$ operations if $n$ is not divisible by $s$), and then implement each group using (i) a QROM look-up of $2^s$ classically computed points $[q]R$ for $q \in [0..2^s - 1]$ and $\lambda_r$-values, (ii) an uncontrolled \EC operation, and (iii) an uncomputation of the QROM, as shown in figure~\ref{fig:window}, thereby reducing the number of calls to point additions. The window size $s$ that minimizes the resource requirements depends on the field size $n$; we discuss this further in section~\ref{sec:QRE}.

\begin{figure}[!ht]
    \centering
    \includegraphics[width=\linewidth]{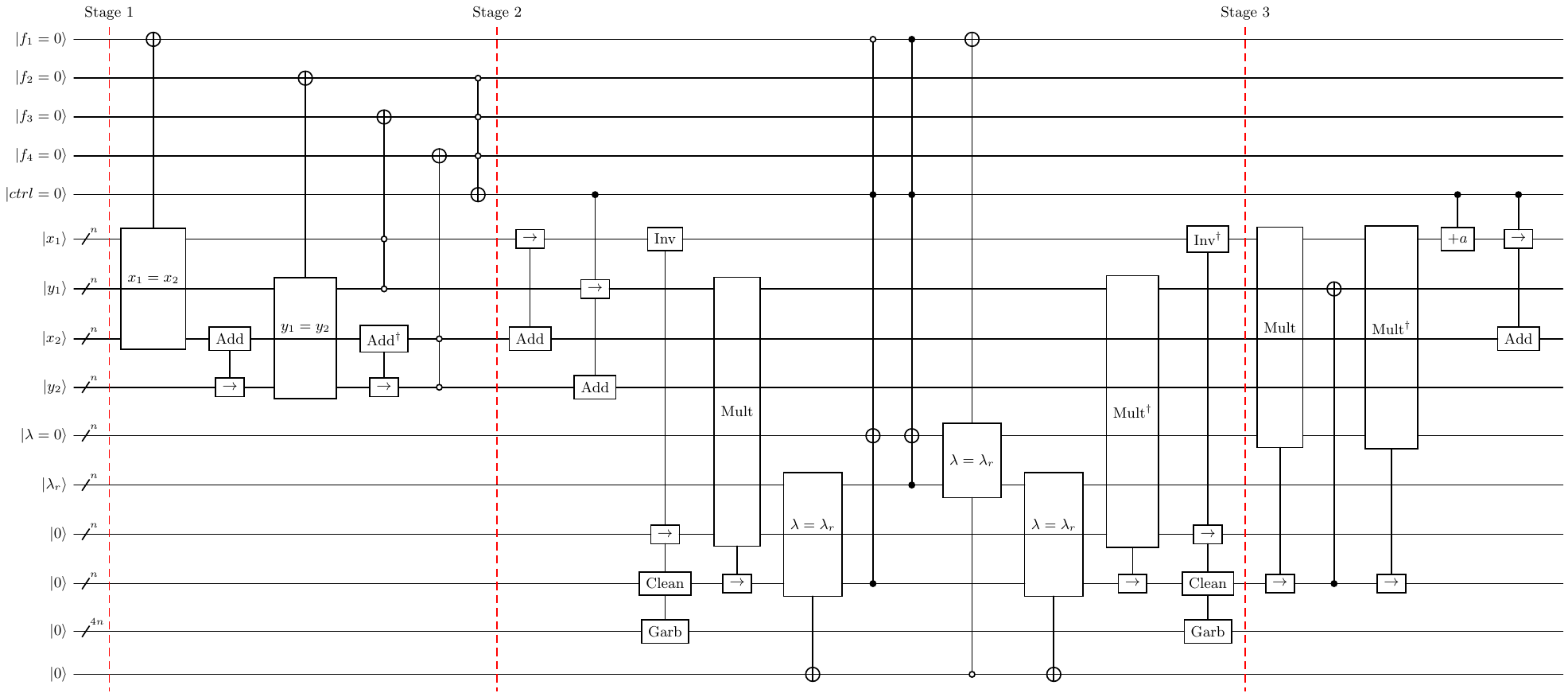}
    \caption{First half of the \EC circuit in figure~\ref{fig:window}. Notations: We use ``$\rightarrow$" to mark the target registers whose values are modified by modular addition, multiplication, and inversion. We use ``Garb" and ``Clean" to label zero-in-garbage-out and zero-in-zero-out ancilla qubits. Note that NOTs with only open controls are Toffoli gates, those with one solid control are series of CNOT gates, and those with one or more solid controls on any of the first five qubits are series of CNOT gates controlled by one or more of the first five qubits.}
    \label{fig:s123}
\end{figure}

\begin{figure}[!ht]
    \centering
    \begin{adjustbox}{width=\textwidth}
    \includegraphics[width=\linewidth]{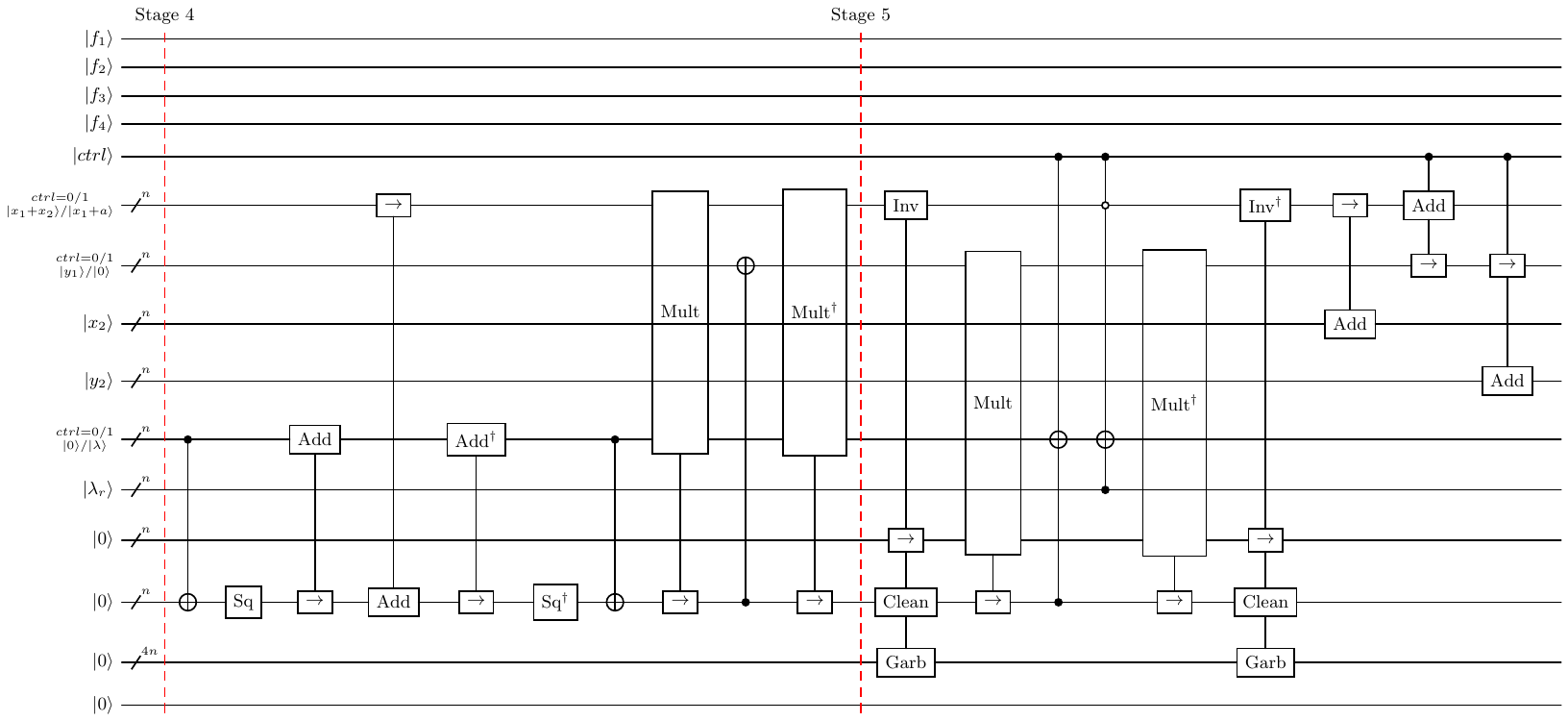}
    \end{adjustbox}
    \begin{adjustbox}{width=0.8\textwidth}
    \includegraphics[width=\linewidth]{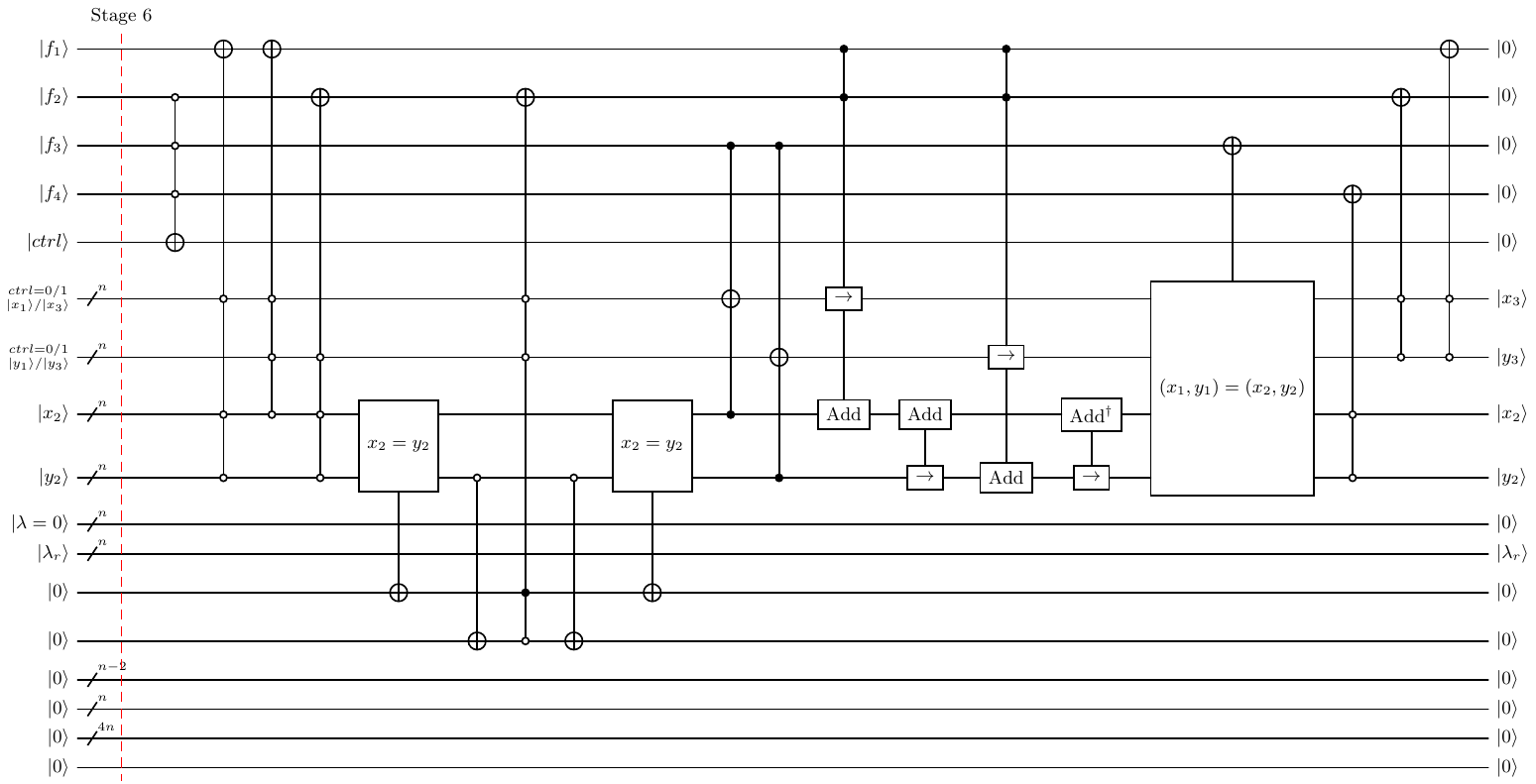}
    \end{adjustbox}
    \caption{Second half of the \EC circuit in figure~\ref{fig:window}. Notations: We denote the squaring operation by ``Sq". In stage 4, the NOT gates controlled by a register are bit-wise CNOT gates. In stages 5 and 6, the controlled NOT gates are all Toffoli gates.}
    \label{fig:s456}
\end{figure}

\subsection{Elliptic curve point addition}
\label{sec:ecptadd}

As a result of our recasting \textsc{ECPointAdd}'s definition, i.e., from~\eqref{eq:point_doubling} and~\eqref{eq:last_case} to~\eqref{eq:combined}, to a form that is similar to the prime \EC definition presented in~\cite{litinski2023compute}, our \EC circuit shares a similar structure to that for prime ECC in~\cite{litinski2023compute}, along with the corrections suggested in~\cite{papa2025validationquantumellipticcurve} and necessary changes -- notably the arithmetic routines -- to adapt to the differences between binary and prime ECC. Unlike previous binary ECC works~\cite{Banegas2020Concrete,Taguchi2023Concrete,Taguchi2024On}, our construction implements the entire point addition including, e.g., the point-doubling case, i.e.,~\eqref{eq:point_doubling}, and the case where either added points is $\mathcal{O}$. Even though the impact of neglecting exceptional cases on the success probability of the algorithm is negligible~\cite{Roetteler2017quantum}, constructing an exact point addition circuit is still of interest~\cite{Larasati2024Quantum}.

Overall, our \EC circuit performs an in-place addition of two points $(x_1,y_1)$ and $(x_2,y_2)$: Each coordinate is stored in an $n$-qubit register, and the output $(x_3,y_3)$ will be returned in the registers that originally stored $x_1$ and $y_1$. Additionally, the circuit requires $n$ qubits to store $\lambda_r$, 5 ancilla qubits to flag the exceptional cases, and $6n$ qubits to implement the conditional logic and arithmetic operations in the point addition. We list and count the subroutines in table~\ref{tb:ecpadd}. In what follows, we explain the circuit in six stages, shown in figures~\ref{fig:s123} and~\ref{fig:s456}.

\begin{table}[!ht]
\begin{tabular}{|l|l|}
\hline
Subroutine                   & Count \\ \hline
$n$-qubit (Controlled) equality test      & 9     \\ \hline
$n$-qubit Toffoli gate       & 30    \\ \hline
Addition                     & 8     \\ \hline
Controlled addition          & 6     \\ \hline
Inversion                    & 4     \\ \hline
Multiplication               & 8     \\ \hline
Controlled constant-addition & 1     \\ \hline
Squaring                     & 2     \\ \hline
\end{tabular}
\caption{Subroutine counts of the \EC circuit in figures~\ref{fig:s123} and~\ref{fig:s456}. In this table, we count a $2n$-qubit Toffoli gate as two $n$-qubit Toffoli gates, since the constant difference in their fault-tolerant costs is negligible. For the same reason, we count a $(n+c)$-qubit Toffoli gate with $c\leq 3$ or $n$ 3-qubit Toffoli gates as a single $n$-qubit Toffoli gate, and a $2n$-qubit equality test as two $n$-qubit ones.}
\label{tb:ecpadd}
\end{table}

In the first stage, we set the flags $f_1, f_2, f_3$, and $f_4$ to 1 when $x_1 = x_2$, $y_1 = x_2 + y_2$, $(x_1,y_1) = \mathcal{O}$, and $(x_2,y_2) = \mathcal{O}$, respectively, using equality checks and multi-controlled Toffoli gates. The $ctrl$ is set to 1 if $f_2=f_3=f_4=0$, meaning that none of the input or output points are $\mathcal{O}$, and that we are in the branch~\eqref{eq:combined} of \textsc{ECPointAdd}. 
In the second stage, we compute $\lambda$ into a clean ancilla register, either (i) using modular arithmetic over binary fields if $f_1=0$ and $ctrl=1$ or (ii) by copying $\lambda_r$ if $f_1=1$ and $ctrl=1$, corresponding to the two cases in~\eqref{eq:combined}. Then, we reset $f_1$ if the outputs of (i) and (ii) are not equal and $\lambda=\lambda_r$, before uncomputing intermediate arithmetic steps. By the end of stage 2, registers 6 and 7 in figure~\ref{fig:s123} are in $\ket{x_1+x_2}$ and $\ket{y_1}/\ket{y_1+y_2}$ if $ctrl=0/1$, and register 10 is in $\ket{0}/\ket{\lambda}$ if $ctrl=0/1$.
Stage 3 maps registers 6 and 7 to $\ket{x_1+x_2}/\ket{x_1+a}$ and $\ket{y_1}/\ket{0}$ respectively, if $ctrl=0/1$; then, stage 4 maps them to $\ket{x_1+x_2}/\ket{x_2+x_3}$ and $\ket{y_1}/\ket{y_2+y_3+x_3}$, respectively, if $ctrl=0/1$. Note that in stage 3, we have used the equality $\lambda=\frac{y_1+y_2}{x_1+x_2}$ from the last line in~\eqref{eq:combined}, and in stage 4, we have used the equalities $x_2 + x_3=\lambda^2+\lambda+x_1+a$ and $\lambda = \frac{y_3+x_3+y_2}{x_2+x_3}$ from~\eqref{eq:combined}.

The fifth stage uncomputes $\ket{\lambda}$, and maps registers 6 and 7 to $\ket{x_1}/\ket{x_3}$ and $\ket{y_1}/\ket{y_3}$ if $ctrl=0/1$, which completes the branch~\eqref{eq:combined} of \textsc{ECPointAdd}. Specifically, this is done by either computing $\lambda$ from $\ket{x_2+x_3}$ (register 6) and $\ket{x_3 + y_2 + y_3}$ (register 7), or by adding $\ket{\lambda_r}$ into $\ket{\lambda}$ when $\ket{x_2+x_3=0}$ and $\ket{ctrl=1}$. $\ket{\lambda_r+\lambda = 0}$ because the conditions $\ket{x_2+x_3=0}$ and $\ket{ctrl=1}$ imply the point-doubling case, i.e., $P_1=P_2$, which we now prove: Along with the conditions, we suppose $P_1\neq P_2$ for the sake of contradiction. Then~\eqref{eq:combined} and $x_2+x_3=0$ imply that $x_1+x_2=0$ because $\lambda = \frac{x_3 + y_2 + y_3}{x_2+x_3} = \frac{y_1+y_2}{x_1+x_2}$. Furthermore, by adding $x_2$ to both sides of $x_3 = \lambda^2 + \lambda + x_1 + x_2 + a$, taken from~\eqref{eq:combined}, and then multiplying both sides by $(x_1+x_2)^2$, we get the equation $(y_1+y_2)^2 + (y_1+y_2)(x_1+x_2) + (x_1+a)(x_1+x_2)=0$. Finally, since $x_1+x_2=0$, the equation simplifies to $(y_1 + y_2)^2 = 0$, which implies $y_1 = y_2$. This leads to a contradiction because if $x_1=x_2$ (implied by $x_1+x_2=0$) and $y+1=y_2$, then $P_1=P_2$. 

In the final stage, we first reset the $ctrl$, and then handle the exceptional cases: The first two Toffoli gates targeting $f_1$ clear $f_1$ for the cases where (i) $(x_2,y_2)=(0,0), x_1 = 0, y_1\neq 0$ and (ii) $(x_1,y_1)=(0,0), x_2 = 0, y_2\neq 0$. The first two Toffoli gates targeting $f_2$, with the latter being conjugated by two Toffoli gates and equality checks, clear $f_2$ for the cases where (i) $(x_2,y_2)=(0,0), x_1 \neq 0, y_1 = 0$ and (ii) $(x_1,y_1)=(0,0), x_2 = y_2 \neq 0$. If $f_3=1$, then $(x_1,y_1)=\mathcal{O}$ and thus, $(x_2,y_2)$ is copied into the output registers, before resetting $f_3$ via an equality check. If $f_4=1$, then $(x_2,y_2)=\mathcal{O}$; thus, we can simply return $(x_1,y_1)$ and reset $f_4$ using a $2n$-qubit Toffoli. If $f_1=f_2=1$ which implies $P_1=-P_2$, then we set the output to $\mathcal{O} = (0,0)$ via arithmetic operations, before resetting the flags.

\subsection{Arithmetic routines}
\label{sec:arith}

Now we proceed to describe the arithmetic operations over binary fields used in our \EC circuit, namely, modular addition, multiplication, and inversion. Note that here we will provide high-level descriptions of these operations and leave the finer details in appendix~\ref{app:arith}.

We first consider modular addition of two polynomials $f(x)$ and $g(x)$, defined by
\begin{equation}
    f(x) + g(x) = \sum_{i=0}^{n-1} (f_i + g_i)x^i,
\end{equation}
where $f_i$ is the $i$th coefficient of $f(x)$ and addition is over $\mathbbm{F}_2$. This is realized by coefficient-wise binary addition, i.e., XOR, which in a quantum circuit, is done by applying $n$ CNOT gates to add $\ket{f}$ into a $n$-qubit input state $\ket{g}$, thereby mapping $\ket{g}$ to $\ket{f+g}$. When $f(x)$ is a classically known constant polynomial, this operation, which we call constant-addition, is realized as an in-place circuit. This circuit consists of at most $n$ NOT gates applied to $\ket{g}$, with one NOT gate for each monomial in $f(x)$ that has a coefficient of one. A controlled addition can be implemented by simply controlling every CNOT, i.e., turning CNOTs into Toffolis. A controlled constant-addition, i.e., $f(x)$ is a constant, is implemented by a series of at most $n$ CNOTs.

%crtmodmult 
Next, we describe modular multiplication between two degree-$(n-1)$ polynomials $f(x)$ and $g(x)$, i.e.,
\begin{equation}
    f(x) \cdot g(x) = f(x) g(x) \bmod{p(x)},
\end{equation}
where $p(x)$ is a degree-$n$ irreducible polynomial. We implement this using the algorithm from~\cite{kim2024toffoli}, with appropriate modifications. This algorithm is based on well-established classical techniques: it combines Karatsuba-like recurrence formulas from~\cite{find2018better,ccalik2019searching} and the Chinese Remainder Theorem (CRT) over binary fields~\cite{sunar2004generalized,fan2007comments} (see theorem 1 in appendix~\ref{appendix:ModMultKKKH}). We choose to use this algorithm because compared to the modular multiplication method from~\cite{van2019space,Banegas2020Concrete}, this algorithm has a much lower Toffoli count.

To start, we note that $f(x) g(x) = f(x) g(x) \bmod{m(x)}$, when $m(x)$ is an arbitrary polynomial with a degree larger than $2n-2$. Furthermore, $m(x)=\prod_{i=1}^t m_i(x)$, where the $m_i(x)$'s are pairwise co-prime polynomials that have degrees $d_i$. We display our choices of $m_i(x)$'s in table~\ref{tab:mod_polys}. Then, using the CRT, we can break the product down into products between smaller polynomials $f^i(x)=f(x) \bmod{m_i(x)}$ and $g^i(x)=g(x) \bmod{m_i(x)}$. The modular reduction to $f^i(x)$ can be computed via $f^i = f_{0,...,d_i-1} + M_i f_{d_i,...,n-1}$, where $f_{i,...,j}$ are a subset of contiguous coefficients, running from $i$ to $j$, of $f(x)$ and $M_i$ is a binary matrix derived from $m_i(x)$. In a quantum circuit, this amounts to XORing certain bits from $\ket{f_{d_i,...,n-1}}$ into $\ket{f_{0,...,d_i-1}}$, where the locations of the CNOTs that perform the XORs are determined by $M_i$~\cite{Banegas2020Concrete}. Next, we compute $c^{i}(x) = f^{i}(x) g^{i}(x) \bmod m_i(x)$ for $i = 1, \ldots, t$. If $d_i \leq 8$, we use existing Karatsuba-like formulas~\cite{find2018better,ccalik2019searching} to compute $c^{i}(x)$; these formulas can be translated to quantum circuits using Algorithm 1 from~\cite{kim2024toffoli} which we explain in appendix~\ref{appendix:ModMultKKKH}. If $d_i>8$, we recursively invoke the CRT-based multiplication algorithm. Following the CRT, we then combine the $c^{i}(x)$'s into a single polynomial using
\begin{equation}\label{eq:combine}
    c^{\prime}(x)=\sum_{i=1}^t \left(c^{i}(x) q_{i}(x)\bmod m(x)\right)\bmod p(x),\: \mbox{where }q_i(x)=\left(\frac{m(x)}{m_i(x)}\right)\left(\left(\frac{m(x)}{m_i(x)}\right)^{-1} \bmod m_i(x)\right),
\end{equation}
where $q_i(x)$'s are classically pre-computed constant polynomials. For each $i$, this involves multiplying $c^{i}(x)$ by the constant polynomial $q_i(x)$, modulo $m(x)$ and $p(x)$; this is a linear transformation of $q_i(x)$ over $\mathbbm{F}_2$ and can be expressed as a matrix-vector multiplication, i.e., $Q_i c^i$, where $Q_i$ is a matrix that depends on $q_i(x)$, $m(x)$, and $p(x)$. Using PLU decompositions, $Q_i$ can be decomposed into a sequence of permutation, lower and upper triangular matrices~\cite{kim2024toffoli}; the permutation and triangular matrices prescribe a sequence of swap and CNOT gates, respectively~\cite{amento2013efficient,Banegas2020Concrete} (see step 3 in appendix~\ref{appendix:ModMultKKKH}). If the degree of $m(x)$ is larger than $2n-2$, then $c^{\prime}(x)$ is the desired product and we are done. Otherwise, i.e., the degree of $m(x)$ $\leq 2n-2$, we need to apply a correction and compute
\begin{equation}\label{eq:correct}
    c(x)=c^{\prime}(x)+\sum_{i=2 n-1-w}^{2 n-2} c_i\left(\left(x^i\right)+\left(x^i \bmod m(x)\right)\right) \bmod p(x),
\end{equation}
where $\omega = 2n - 1 - deg(m(x))$, and $c_i$'s are referred to as correction coefficients. To implement this in a quantum circuit, the multiplication by the classically pre-computed polynomial $\left(\left(x^i\right)+\left(x^i \bmod m(x)\right)\right) \bmod p(x)$ is carried out using the previously mentioned PLU decomposition method. We notice that the circuit provided in~\cite{kim2024toffoli} for computing $c_i$'s is incorrect. We correct the circuit by applying appropriate CNOTs and without incurring additional Toffolis, as shown in figure~\ref{fig:HighDegreeCircuit}, and subsequently, roughly halving its CNOT count. See step 4 in appendix~\ref{appendix:ModMultKKKH} for details. Note that this circuit requires both CNOTs and Toffolis, and it is much smaller compared to the previous parts of the modular multiplication algorithm. 

The modular inversion operation computes the inverse of a given polynomial $f(x)$ modulo ${p(x)}$, denoted as 
\begin{equation}
    f^{-1}(x) \bmod{p(x)}.
\end{equation}
Using an extension of Fermat's Little Theorem (FLT) over binary fields, the modular inverse can be equivalently obtained by computing~\cite{itoh1988fast}
\begin{equation}
f(x)^{2^n - 2} = f^{-1}(x) \bmod{p(x)}.
\end{equation}
Moreover, $f(x)^{2^n - 2}$ can be computed using a sequence of squaring and modular multiplication operations on appropriate powers of $f(x)$. A quantum circuit implementation of this FLT-based inversion method was first described in~\cite{amento2012quantum}. This FLT-based approach was compared to a greatest common divisor (GCD)-based approach in~\cite{Banegas2020Concrete}, which found that the former has a lower Toffoli count while the latter has a lower qubit count. To address this trade-off, we make use of the FLT-based algorithm introduced in~\cite{Taguchi2024On}, which improves on previous FLT-based algorithms to reduce the number of ancilla qubits. For a comparison, see appendix~\ref{appendix:InversionTT24} for a discussion about a more Toffoli-optimal FLT-based algorithm from~\cite{Taguchi2023Concrete}. This algorithm is based on the observation that in FLT-based algorithms, when computing $f(x)^{2^n - 2}$, the exponents of $f(x)$ form an addition chain~\footnote{An addition chain for a non-negative integer $n-1$ is a sequence $\alpha_0 = 1, \alpha_1 , \alpha_2, \ldots, \alpha_l = n-1$, with the property that each $\alpha_i$, after $\alpha_0$, is obtained by adding two earlier terms that are not necessarily distinct.} for $2^n - 2$. These addition chains correspond to distinct sequences of squaring and modular multiplication operations. Furthermore, one can find alternative addition chains and thus, sequences of squaring and modular multiplication that optimize the number of intermediate terms, i.e., powers of $f(x)$, which can be uncomputed to $\ket{0...0}$ ancilla qubits that are reused throughout the algorithm, thereby reducing the number of ancilla qubits required. However, this spatial optimization comes at the cost of an increase in the number of squaring and modular multiplication operations. Notably, compared to the GCD-based method in~\cite{Banegas2020Concrete} over relevant field sizes, the Toffoli counts remain much lower and the qubit counts are competitive.

For modular multiplication, we use the CRT-based method described earlier. The modular squaring of a polynomial $f(x)$, i.e., $f(x)^2\bmod{p(x)}$, can be formulated as a multiplication between a matrix $S$ on the vector coefficients of $f(x)$, i.e., $\ket{f} \mapsto \ket{Sf}$. In particular, $S$ combines two actions: (i) it maps the monomials $f_i x^{i}$ to $f_i x^{2i}$, and (ii) it performs modular reduction with respect to $p(x)$; see appendix~\ref{appendix:sq}. In the quantum circuit, this combined operation can be realized as a sequence of swap and CNOT gates, using the previously mentioned PLU decomposition method~\cite{amento2013efficient}.

\section{Resource Estimation}
\label{sec:QRE}
In this section, we report our resource estimation methodologies and the resulting estimates of our algorithm applied to binary ECC of cryptographically relevant field sizes: $n \in \{ 163,233,283,571 \}$~\cite{kerry2013digital}. We begin by describing the quantum circuits at the logical level, accounting for both Clifford and non-Clifford gates; both types of gates are necessary to estimate the active volume (AV) of the algorithm~\cite{litinski2022activevolume}, whereas only the non-Clifford gate count is needed to estimate the circuit volume of the algorithm~\cite{Litinski2019gameofsurfacecodes,litinski2022activevolume}. From there, we estimate the hardware footprint of and runtime on baseline and AV architectures.

\begin{table}[!ht]
\begin{tabular}{|l|l|l|l|l|l|}
\hline
$n$ & $\#$ CNOTs   & $\#$ Swaps & $\#$ Toffolis & Active Volume     \\ \hline
163 & 110956  & 300   & 999        & $4.91 \times 10^5$  \\ \hline
233 & 225402  & 448   & 1448       & $9.70 \times 10^5$  \\ \hline
283 & 325206  & 618   & 1776       & $1.38 \times 10^6$ \\ \hline
571 & 1287610 & 2208  & 3860       & $5.33 \times 10^6$ \\ \hline
\end{tabular}
\caption{The costs of the CRT-based modular multiplication algorithm stated in terms of CNOTs, swaps, Toffolis, and active volume. Note that $2n$ input qubits store polynomials $f(x)$ and $g(x)$, and $n$ output qubits store the result $h(x) + f(x) \cdot g(x) \bmod{p(x)}$.}
\label{tab:modmult_results}
\end{table}

%% INVERSION TT24
\begin{table}[h!]
\begin{tabular}{|l|l|l|l|l|l|l|}
\hline
$n$   & $\#$ ModMults  & $\#$ CNOTs & $\#$ Swaps &  $\#$ Toffolis & Active Volume \\ \hline
163 & 14                     & 1651326     & 14765       & 13986             & $7.26 \times 10^6$   \\ \hline
233 & 16                     & 3761228     & 55298       & 23168             & $1.61 \times 10^7$  \\ \hline
283 & 18                     & 6254129     & 47997       & 31968             & $2.65 \times 10^7$  \\ \hline
571 & 20                     & 27646645    & 134422      & 77200             & $1.14 \times 10^8$ \\ \hline
\end{tabular}
\caption{The costs of computing $ f^{-1}(x)\bmod{p(x)}$ given $f(x)$ via a FLT-based inversion algorithm~\cite{Taguchi2024On} stated in terms of the number of modular multiplication applications, CNOTs, swaps, and Toffolis, as well as active volume. $2n$ qubits are used to store the input and output, and for the field sizes $n$ stated here, the number of ancilla qubits is $5n$.}
\label{tab:inversion_results}
\end{table}

\begin{table}[!ht]
\begin{tabular}{|l|l|l|l|}
\hline
$n$   & $\#$ Toffolis  & $\#$ Qubits & Active Volume \\ \hline
163 & $7.13 \times 10^4$           & 1963 & $3.33 \times 10^7$         \\ \hline
233 & $1.15 \times 10^5$           & 2803   & $7.29 \times 10^7$      \\ \hline
283 & $1.55 \times 10^5$         & 3403     & $1.18 \times 10^8$     \\ \hline
571 & $3.65 \times 10^5$         & 6859   & $5.01 \times 10^8$      \\ \hline
\end{tabular}
\caption{The costs of an \EC circuit.}
\label{tab:point_doubling_results}
\end{table}

\begin{table}[!ht]
\begin{tabular}{|l|l|l|l|l|}
\hline
$n$ & $s$ & $\#$ Toffolis  & Qubits    & Active Volume     \\ \hline

163 & 13        &  $2.05 \times 10^6$           & 2126 &   $9.50 \times 10^8$ \\ \hline
233 & $13/14$       &     $4.42 \times 10^6$              & 3036 &   $ 2.78\times 10^9$\\ \hline
283 & 15       &  $7.09\times 10^6$            & 3686&    $5.30\times 10^9$ \\ \hline
571 & 16       &    $3.09\times 10^7$            & 7430 &   $ 4.22\times 10^{10}$ \\ \hline
\end{tabular}
\caption{The costs of the phase estimation circuit in figure~\ref{fig:Shor} implemented using the windowing circuit in figure~\ref{fig:window}. We optimize the window size $s$ for the number of Toffolis and active volume separately. Note that the optimal $s$-values for Toffolis and active volume are not necessarily the same but incidentally, they are the same for all considered field sizes except $n=233$, for which the optimal $s$-values for Toffolis and active volume are 13 and 14, respectively.}
\label{tab:phase_estimation_results}
\end{table}

\begin{table}[!ht]
\begin{tabular}{|l|l|l|l|l|}
\hline
$n$ & $s$ & $\#$ Toffolis  & Qubits    & Active Volume     \\ \hline

163 & 13        &  $1.42 \times 10^6$           & 2126 &   $6.43\times 10^8$ \\ \hline
233 & 14       &     $ 3.65 \times 10^6$              & 3036 &   $ 2.24\times 10^9$\\ \hline
283 & 14       &  $ 5.80 \times 10^6$            & 3686&    $4.31 \times 10^9$ \\ \hline
571 & 15       &    $2.78 \times 10^7 $            & 7430 &   $ 3.78 \times 10^{10}$ \\ \hline
\end{tabular}
\caption{The costs of the phase estimation circuit in figure~\ref{fig:Shor} implemented using the windowing circuit in figure~\ref{fig:window} and assuming $48$ classically pre-computed bits. We optimize the window size $s$ for the number of Toffolis and active volume separately.}
\label{tab:phase_estimation_results_2}
\end{table}

\begin{figure}[!ht]
\centering
\begin{subfigure}{0.5\textwidth}
  \centering
  \includegraphics[width=0.765\linewidth]{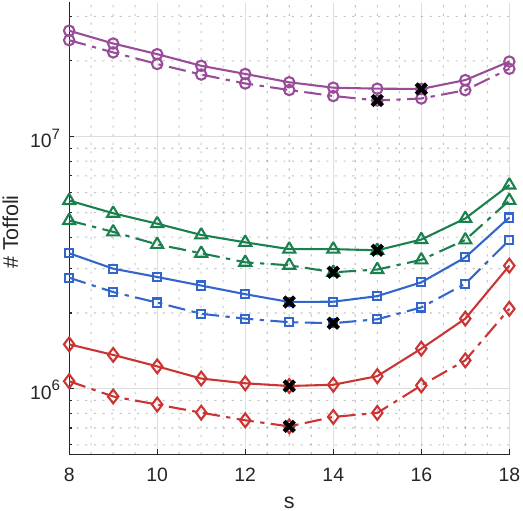}
  \caption{The cost landscape with respect to Toffoli count.}
  \label{fig:toff_landscape}
\end{subfigure}%
\begin{subfigure}{.5\textwidth}
  \centering
  \includegraphics[width=\linewidth]{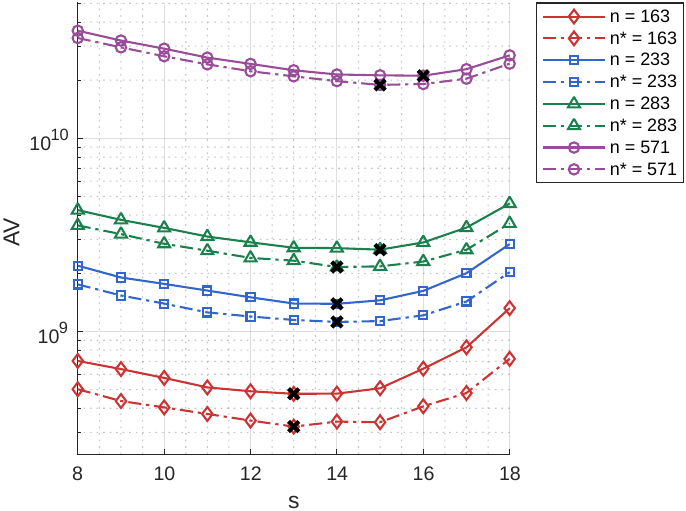}
  \caption{The cost landscape with respect to active volume.}
  \label{fig:av_landscape}
\end{subfigure}
\caption{The cost landscapes of a single round of phase estimation, i.e., half of the circuit in figure~\ref{fig:Shor}, plotted against window size $s$. The optimal window sizes with respect to Toffoli count and active volume are marked by crosses in (a) and (b), respectively. (a) and (b) share the same legend, in which the asterisks mark the cases with 48 classically precomputed bits in the private key.}
\label{fig:cost_landscape}
\end{figure}

\subsection{Estimating gate and qubit counts} 

First, we estimate the gate and qubit counts of the three main circuit components in our algorithm, namely, the \EC circuit, QROM look-up and its uncomputation, disregarding the comparatively negligible cost of quantum Fourier transform as done in~\cite{Banegas2020Concrete,litinski2022activevolume,Taguchi2023Concrete,Taguchi2024On}.

Starting with the non-arithmetic components in \textsc{ECPointAdd}, we implement each $n$-qubit equality check as an $n$-qubit Toffoli conjugated by $n$ pairs of bitwise CNOT gates, and each $n$-qubit Toffoli is decomposed into $n-1$ regular Toffoli gates and $n-1$ clean ancilla qubits~\cite{PhysRevA.87.022328}. 
Next, we summarize the arithmetic circuits, which are explained in section~\ref{sec:arith} and appendix~\ref{app:arith}. A (controlled) modular addition is implemented by $n$ bit-wise (controlled) CNOTs applied to the addends. A controlled modular constant-addition, which is only used once per \EC circuit, can be implemented using at most $n$ CNOTs, which we round up to $n$ CNOTs in our resource estimates. The CRT-based modular multiplication is implemented using circuits that are constructed from pre-computed polynomials $m_i(x)$'s and $p(x)$, the Karatsuba-like multiplication circuits for low-degree polynomials from~\cite{kim2024toffoli}, and the correction circuit in figure~\ref{fig:HighDegreeCircuit}. In particular, we generate the CNOT circuits for modular reductions from $m_i(x)$'s, and the circuits, consisting of only swap and CNOT gates, for modular multiplication with pre-computed constant polynomials, used in~\eqref{eq:combine} and~\eqref{eq:correct}, from $m_i(x)$'s and $p(x)$. Note that the Karatsuba-like multiplication circuit incurs the majority of the Toffoli gates in the modular multiplication algorithm, with a minor contribution from the correction circuit if it is applied. The FLT-based modular inversion circuit comprises repeated calls to modular multiplication and squaring, which is a circuit built from CNOTs and swaps. We provide \emph{exact} gate and qubit counts -- computed from the classical inputs of the arithmetic circuits -- for an application of modular multiplication in table~\ref{tab:modmult_results} and those for an application of modular inversion in table~\ref{tab:inversion_results}. The resource estimates of a fully compiled \EC circuit are listed in table~\ref{tab:point_doubling_results}.

We choose to use the QROM implementation from~\cite{PhysRevX.8.041015} and its clean-ancilla-assisted uncomputation from~\cite{Berry2019qubitizationof} due to their low Toffoli counts; a $k$-item QROM look-up costs $k-2$ Toffoli gates and the uncomputation circuit costs approximately $2\sqrt{k}$ Toffoli gates. Then, the total Toffoli count of the phase estimation circuit in figure~\ref{fig:Shor} is two times the following:
\begin{equation}
    \left\lfloor\frac{n}{s}\right\rfloor \left(2^s-2 + \mathcal{C}(\textsc{ECPointAdd}) + 2^{s/2 + 1}\right) + \left(2^{n \bmod{s}} -2 + \mathcal{C}(\textsc{ECPointAdd}) + 2^{(n \bmod{s})/2 + 1}\right),
\end{equation}
where $\mathcal{C}(\textsc{ECPointAdd})$ is the cost of \EC and $s$ is the window size of the circuit in figure~\ref{fig:window}. We minimize the Toffoli count over $s$, and list the optimized Toffoli count and qubit count, including the ancilla qubits used to synthesize the multiply controlled Toffolis, in table~\ref{tab:phase_estimation_results}. We plot the Toffoli count at various $s$-values in figure~\ref{fig:toff_landscape}. Note that our Toffoli counts are lower than those in~\cite{Taguchi2024On}, where the same arithmetic routines are used. 

As in~\cite{litinski2023compute}, we also consider the scenario where 48 bits of the private key are determined on a classical computer using algorithms in~\cite{cryptoeprint:2015/605}, which require at most $2^{25}$ point addition operations. According to~\cite{koblitz2000state}, a million-instruction-per-second, i.e., MHz, CPU can perform 40000 point additions per second. Therefore, computing the 48 bits will take about 14 minutes on a MHz CPU, and a thousandth of that, i.e., less than a second, on a GHz CPU, which is negligible compared to the runtime on the quantum computer, as we discuss below. The Toffoli counts to compute the remaining $n-48$ bits in this case are listed in table~\ref{tab:phase_estimation_results_2} and plotted in figure~\ref{fig:toff_landscape}.

\subsection{Estimating hardware footprint and runtime} 

In what follows, we delineate our methodology for translating the logical circuits discussed above to estimates of the hardware footprint and time required to execute our algorithm. We consider (i) a baseline architecture executed on a superconducting or trapped-ion quantum computer~\cite{Litinski2019gameofsurfacecodes}, and (ii) an AV architecture~\cite{litinski2022activevolume} executed on a photonic FBQC~\cite{bartolucci2023fusion}. These architectures implement surface codes~\cite{KITAEV20032,bravyi1998quantumcodeslatticeboundary} and execute logical operations via lattice surgery~\cite{horsman2012surface,Litinski2019gameofsurfacecodes}.

In both architectures, the footprint in space and time of a computation is determined by a quantity called \emph{spacetime volume}, which in turn, depends on the number of logical qubits and logical cycles required to execute the computation. A logical cycle comprises $d$ code cycles, where $d$ is the code distance and a code cycle is the time needed to perform a syndrome measurement. Since a lattice surgery operation is implemented in a logical cycle, we measure computational time in units of logical cycles. We now proceed to quantify the spacetime volume for both architectures, following the methods in~\cite{litinski2022activevolume,caesura2025fasterquantumchemistrysimulations}.

In the baseline architecture, a logical circuit's spacetime volume is roughly twice its circuit volume $n_Q\cdot n_T$, where $n_Q$ and $n_T$ are the number of qubits and T gates, respectively, in the circuit. We take $n_T$ to be four times the number of Toffolis because a Toffoli can be decomposed into four T gates~\cite{PhysRevA.87.022328}. 
Each T gate is implemented via multi-qubit Pauli measurements between a number of qubits and a magic state. 
This architecture assumes roughly $2n_Q$ qubits laid out on a two-dimensional grid: $n_Q$ qubits, which we call memory qubits, are from the abstract logical circuit, $n_W = n_Q$ qubits, which we call workspace qubits, are used to mediate Pauli measurements between memory qubits, and a much smaller group of qubits is reserved for distilling T gates to be consumed by the workspace qubits. It is further assumed that one T gates is produced per logical cycle, and that the T gate is consumed by workspace qubits while the next T gate is being produced. As a result, the total runtime is then proportional to $n_T$, and the spacetime volume is roughly given by 
\begin{equation}
    V_B = 2 n_Q\cdot n_T.
\end{equation}

In the AV architecture, the spacetime volume scales with the circuit's AV instead of circuit volume. AV measures the number of lattice surgery~\cite{horsman2012surface} operations used to execute a circuit, while leveraging the parallelism made possible by the fact that each surface-code qubit connected to logarithmically many qubits~\cite{litinski2022activevolume}. There are a total of $n_Q+n_W$ qubits, where $n_Q$ and $n_W$ are the number of memory and workspace qubits, respectively; both logical operations and magic-state distillation are performed using the workspace qubits. The set of logical, lattice surgery operations utilized in the AV architecture are known as logical blocks~\cite{PRXQuantum.4.020303,litinski2022activevolume}. The AV architecture further assumes the execution of $n_W$ logical blocks per logical cycle.  We count AV in the number of logical blocks. To estimate the AV of a circuit, one could express the circuit in terms of elementary operations, the AV of which are listed in table 1 from~\cite{litinski2022activevolume}, and sum up the AV of the circuit's constituents. We have done that for the arithmetic circuits in our algorithm and listed their AV in tables~\ref{tab:modmult_results}-\ref{tab:phase_estimation_results}. For the non-arithmetic components, we take the AV estimates from figure 3 in~\cite{litinski2023compute}, except for the $k$-item QROM uncomputation circuit. Our QROM uncomputation circuit, taken from~\cite{Berry2019qubitizationof}, has an AV of $\approx 0.75k + 120 \sqrt{k}$, which is lower than that of the uncomputation circuit used in~\cite{litinski2023compute}, because of the use of a Toffoli-free measurement uncomputation of a binary-to-unary circuit. Note that we minimize the window sizes $s$ with respect to AV, separately from Toffoli count, and list the optimal $s$-values, with and without classical precomputed bits in the private key, in tables~\ref{tab:phase_estimation_results} and~\ref{tab:phase_estimation_results_2}, respectively. We plot the AV at various $s$-values in figure~\ref{fig:toff_landscape}. A circuit with a total AV of $b_{AV}$ has an estimated spacetime volume of
\begin{equation}
    V_A = \frac{b_{AV}}{n_W} (n_Q + n_W).
\end{equation}
Hereafter, we set $n_W = n_Q$ and thus $V_A = 2 b_{AV}$, as in~\cite{litinski2022activevolume,litinski2023compute}. Although, in general, the ratio between logical and workspace qubits can be adjusted to tune the computational performance, as demonstrated in~\cite{caesura2025fasterquantumchemistrysimulations}.

Now we proceed to calculate the physical resources, i.e., hardware footprint and runtime, it takes to execute a circuit. We first consider the baseline architecture on a matter-based device: The execution time of a circuit can be estimated as
\begin{equation}
    \mathcal{T} = \mbox{number of code cycles} \times \mbox{code cycle time} = d \times \mbox{number of logical cycles} \times \mbox{code cycle time},
\end{equation}
where $d$ is the code distance. We assume a logical error rate per unit of spacetime volume to be $p_L = 10^{-d/2}$, corresponding to 10$\%$ of the surface code threshold~\cite{litinski2023compute,PRXQuantum.4.020303}, and a $0.05$ failure probability for a single computation, as in~\cite{litinski2023compute}. Then, $d$ can be obtained by solving
\begin{equation}\label{eq:error_rate}
    p_L\cdot V_B \leq 0.05.
\end{equation}
Note that this failure probability can also be construed as a failure probability $\leq 0.5$ after 10 executions of the algorithm, i.e., one in 10 executions is faulty. So, the average runtime of a successful computation will be $10\mathcal{T}/9$. The number of logical cycles is given by $n_T$, and the code cycle time depends on the hardware. We assume a hardware-motivated 1 $\mu$s and 1 ms code cycle time for a hypothetical superconducting and trapped-ion (or neutral-atom) device, respectively, in line with~\cite{litinski2022activevolume,Gidney2021how,litinski2023compute,viszlai2023architectureimprovedsurfacecode,xu2024constant,beverland2022assessingrequirementsscalepractical}. The number of physical qubits can be estimated as $2 n_Q \cdot d^2$, where $d^2$ is the number of physical qubits per surface-code logical qubit. We list these physical resource estimates, including the number of physical qubits and average runtime, for our algorithm, with and without classically precomputed bits in the private key, in table~\ref{tab:bl_qre}. 

We move onto the AV architecture on a photonic FBQC, where computation is carried out by performing multi-photon entangling measurements called fusions on photonic entangled resource states~\cite{bartolucci2023fusion}. We measure the physical footprint of the computer in the number of physical units called \emph{interleaving modules} (IMs)~\cite{bombin2021interleavingmodulararchitecturesfaulttolerant,litinski2022activevolume}. Each IM comprises a collection of resource state generators (RSGs) and delay lines that store and route the generated resource states to fusion locations~\cite{bombin2021interleavingmodulararchitecturesfaulttolerant,litinski2022activevolume}. The execution time $\mathcal{T}$ of a circuit is given by
\begin{equation}
    \mathcal{T} = \frac{V_A d^3}{n_{IM} r_{IM}},
\end{equation}
where the numerator is the spacetime volume given in the units of resource states with $d^3$ being the number of resource state per logical block, $n_{IM}$ is the number of IMs, and $r_{IM}$ is the total resource state generation rate per IM, i.e., the sum of the rate of its constituent RSGs. We compute $d$ by substituting $V_B$ for $V_A$ in~\eqref{eq:error_rate} and then solving it. As done for the baseline architecture, we take $10\mathcal{T}/9$ to be the average runtime of a successful computation. Moreover, we take $r_{IM}$ to be 1GHz as in~\cite{litinski2022activevolume,caesura2025fasterquantumchemistrysimulations}; note that $r_{IM}$ is unrelated to the physical clock rate, since an IM with $x$ RSGs at a rate $y$ has the same $r_{IM} = xy$ as one with $a\cdot x$ RSGs at a rate $y/a$. $n_{IM}$ can be computed as follows:
\begin{equation}\label{eq:nim}
    n_{IM} = \frac{n d^2}{n_{RS/IM}},
\end{equation}
where $n d^2$ is the number of resource states needed to construct $n=n_Q+n_W$ logical qubits, assuming an architecture based on six-ring resource states~\cite{bartolucci2023fusion}, and $n_{RS/IM}$ is the number of resource states stored in an IM at any given time, defined by
\begin{equation}\label{eq:rsim}
    n_{RS/IM} = r_{IM}\cdot \frac{l_{delay}}{c_{fiber}},
\end{equation}
where $l_{delay}$ is the length of a fiber optic delay line and $c_{fiber} = 2 \times 10^8 m/s$ is the speed of light in a fiber optic cable. Combining~\eqref{eq:nim} and~\eqref{eq:rsim}, we obtain
\begin{equation}
    n_{IM} = \frac{nd^2 c_{fiber}}{r_{IM} l_{delay}}.
\end{equation}
We list these physical resource estimates, including the number of IMs and average runtime, for our algorithm, with and without classically precomputed bits in the private key, in table~\ref{tab:av_qre}.

\begin{table}[!ht]
\begin{tabular}{|l|l|l|l|l|}
\hline
Field size $n$                                    & 163                & 233                & 283                  & 571                  \\ \hline
Distance $d$                                     & 24                 & 25                 & 26                   & 28                   \\ \hline
Device size                         & $2.45 \times 10^6$ & $3.79 \times 10^6$ & $4.98\times 10^6$    & $1.16\times 10^7$    \\ \hline
Runtime $\Big(\substack{\mbox{$1\mu$s/ 1ms}\\ \mbox{code cycle}}\Big)$ & $\substack{\mbox{3.6 min /} \\ \mbox{2.5 days}}$ & $\substack{\mbox{8.2 min /} \\ \mbox{5.7 days}}$ & $\substack{\mbox{13.7 min /} \\ \mbox{9.5 days}}$ & $\substack{\mbox{64.1 min /} \\ \mbox{44.5 days}}$ \\ \hline
$\mbox{Runtime}^{\ast}$ $\Big(\substack{\mbox{$1\mu$s/ 1ms}\\ \mbox{code cycle}}\Big)$ & $\substack{\mbox{2.5 min /} \\ \mbox{1.8 days}}$ & $\substack{\mbox{6.8 min /} \\ \mbox{4.7 days}}$ & $\substack{\mbox{11.2 min /} \\ \mbox{7.8 days}}$ & $\substack{\mbox{57.7 min /} \\ \mbox{40.0 days}}$ \\ \hline
\end{tabular}
\caption{The physical resource estimates for baseline architectures executed on a trapped-ion or neutral-atom device with a 1ms code cycle and a superconducting device with a $1\mu$s code cycle. The device size is measured in the number of physical qubits. In the asterisked row, we list the runtimes with 48 classically precomputed bits in the private key.}
\label{tab:bl_qre}
\end{table}

\begin{table}[!ht]
\begin{tabular}{|l|l|l|l|l|}
\hline
Field size $n$                                    & 163                & 233                & 283                  & 571                  \\ \hline
Distance $d$                                     & 22                 & 23                 & 23                   & 25                   \\ \hline
Device size $\Big(\substack{\mbox{$1\mu$s delay} / \\ \mbox{$10\mu$s delay}}\Big)$  &   $\substack{  \mbox{2058} / \\ \mbox{206}}$ & $\substack{ \mbox{3213} / \\ \mbox{322}}$ & $\substack{ \mbox{3900} /  \\ \mbox{390} }$ &   $\substack{ \mbox{9288} /  \\ \mbox{929}}$  \\ \hline
Runtime $\Big(\substack{\mbox{$1\mu$s delay} / \\ \mbox{$10\mu$s delay}}\Big)$ & $\substack{\mbox{10.9 sec/} \\ \mbox{1.8 min}}$ & $\substack{\mbox{23.4 sec /} \\ \mbox{3.9 min}}$ & $\substack{\mbox{36.7 sec /} \\ \mbox{6.1 min}}$ & $\substack{\mbox{2.6 min /} \\ \mbox{26.3 min}}$ \\ \hline
Distance $d^{\ast}$                                     & 21                 & 22                 & 23                   & 25                   \\ \hline
$\mbox{Device size}^{\ast}$ $\Big(\substack{\mbox{$1\mu$s delay} / \\ \mbox{$10\mu$s delay}}\Big)$  &   $\substack{\mbox{1876} / \\ \mbox{188}}$ & $\substack{ \mbox{2939} / \\ \mbox{294}}$ & $\substack{ \mbox{3900} /  \\ \mbox{390} }$ &   $\substack{ \mbox{9288} /  \\ \mbox{929}}$  \\ \hline
$\mbox{Runtime}^{\ast}$ $\Big(\substack{\mbox{$1\mu$s delay} / \\ \mbox{$10\mu$s delay}}\Big)$ & $\substack{\mbox{7.0 sec/} \\ \mbox{1.2 min}}$ & $\substack{\mbox{18.0 sec /} \\ \mbox{3.0 min}}$ & $\substack{\mbox{29.9 sec /} \\ \mbox{4.9 min}}$ & $\substack{\mbox{2.4 min /} \\ \mbox{23.5 min}}$ \\ \hline
\end{tabular}
\caption{The physical resource estimates for active architectures executed on a photonic FBQC with $1\mu$s and $10\mu$s delay. The device size is measured in the number of interleaving modules with a 1 GHz total resource state generation rate. In the asterisked rows, we list the resource estimates with 48 classically precomputed bits in the private key.}
\label{tab:av_qre}
\end{table}

\section{Discussion}
\label{sec:discussion}
We estimated both the logical and physical resource costs of computing binary elliptic curve discrete logarithms on a fault-tolerant quantum computer. In addition, we constructed the first, to our knowledge, quantum circuit that implements the point addition operation exactly, including all exceptional cases. We also corrected and optimized the quantum circuit implementation~\cite{kim2024toffoli} of the CRT-based modular multiplication algorithm~\cite{sunar2004generalized,fan2007comments}. Compared to prior art~\cite{Taguchi2024On} that uses the same arithmetic algorithms as we do, albeit with the uncorrected modular multiplication circuit, our algorithm incurs a lower Toffoli count over cryptographically relevant binary field sizes, due to the use of a more efficient QROM uncomputation circuit.

We carried out the physical resource estimation, i.e., hardware footprint and runtime, for a baseline architecture executed on hypothetical superconducting, trapped-ion, and neutral-atom quantum computers, and for the active volume architecture executed on a hypothetical photonic fusion-based quantum computer. Comparing to the superconducting baseline architecture, the photonic active volume architecture executes our algorithm over 20 times faster assuming a $1 \mu s$ delay and 2 times faster assuming a $10 \mu s$ delay, over cryptographically relevant field sizes. The speed-ups can be understood from perspectives discussed in~\cite{litinski2022activevolume}: (i) In a baseline architecture, the spacetime cost per Toffoli is proportional to the number of logical qubits, while it is independent of the number of logical qubits in the active volume architecture. (ii) The reduced code distance on an active-volume device offers another source of speed-up. (iii) The length of delay acts as a trade-off parameter between speed and hardware footprint, i.e., number of resource state generators. On a trapped-ion or neutral-atom device, the speed-ups are magnified by a factor of 1000 due to its slower logical clock-speed. 

Compared to the algorithms for computing prime elliptic curve discrete logarithms~\cite{litinski2023compute} and factoring a 2048-bit RSA integer~\cite{Gidney2021how}, our algorithm has a much lower Toffoli count -- one to two orders of magnitude depending on field size -- and a faster runtime on both baseline and active volume architectures~\cite{litinski2022activevolume}. The hardware footprint and qubit count of our algorithm are similar to those of the prime-curve algorithm for comparable field sizes, and are smaller than those of the 2048-RSA algorithm. 

However, the AV-to-Toffoli-count ratio of our algorithm is much higher compared to that of the prime ECC algorithm in~\cite{litinski2023compute}, which leads to a relatively smaller speed-up from using the AV architecture. This is due to the large number of CNOTs in the modular multiplication circuit. Compared to a close alternative -- the Karatsuba multiplication circuit in~\cite{van2019space}, our chosen circuit's Toffoli count is about an order of magnitude smaller, but its CNOT count is higher by about $2-4$ times over the considered field sizes; while our chosen circuit still has a lower AV, there could be more AV-optimal multiplication circuits, which we leave for investigation in future work. An alternative way to improve our AV estimates is via peephole optimization. Instead of simply adding up the AV in a gate-by-gate manner, since AV is subadditive~\cite{litinski2022activevolume}, we could optimize the AV of subcircuits, which consist of a collection of gates, before adding them up. For example, the modular multiplication circuit comprises many purely CNOT subcircuits, over which one could perform gate optimization using methods in, e.g.,~\cite{doi:10.1137/1.9781611975994.13,9792395,8335339,patel2008}, which in turn reduces AV, or perform direct AV optimization using tools like ZX calculus~\cite{litinski2022activevolume}. Such optimizations are better suited to be carried out via software. We leave such explorations for future work.

In addition, we have left out a couple of potential optimizations: (1) The modular division operation can be parallelized over multiple instances of the algorithm~\cite{litinski2023compute}. (2) By tuning the ratio between workspace and memory qubits, the performance of both architectures can be enhanced, see, e.g.,~\cite{10.1145/3689826,caesura2025fasterquantumchemistrysimulations,Gidney2021how}. We further provide some interesting directions to be pursued in the future: (1) Perform a similar study of the recently invented Regev's algorithm~\cite{10.1145/3708471,10.1007/978-3-031-62746-0_10} applied to binary elliptic curve discrete logarithms, and compare it with our algorithm. (2) Extend Karatsuba-like formulas~\cite{find2018better,ccalik2019searching} to $10$-way splits. Then, we will not have to recursively call CRT-based modular multiplication for larger field sizes, which in turn, will lead to lower gate counts and AV.

\section*{Acknowledgements}
A. K. thanks Sukin Sim for insightful discussions on physical resource estimation methodologies, and Daniel Litinski and Sam Pallister for valuable comments on the manuscript.

\nocite{}

\bibliography{ref}
\bibliographystyle{apsrev4-2}

\appendix

\section{Details on arithmetic subroutines}
\label{app:arith} 
In this section, we present the essential subroutines and circuits required for the elliptic curve point addition circuit in section~\ref{sec:algo}. These subroutines can be constructed from Toffoli, CNOT, and swap gates. The structure of this section is as follows: section~\ref{appendix:toff_free} provides Toffoli-free modular arithmetic circuits, section~\ref{appendix:ModMultKKKH} describes how to perform modular multiplication (the primary contributor to the Toffoli gate count), and section~\ref{appendix:InversionTT24} describes the modular inversion subroutine.

\subsection{Toffoli-free arithmetic}
\label{appendix:toff_free}
In this section, we summarize arithmetic operations over binary fields that do not require Toffoli gates. We begin by defining basic notation and then describe several subroutines necessary for modular multiplication and inversion algorithms. These subroutines are categorized into two types: out-of-place and in-place algorithms. The subroutine circuits are determined by classical inputs, with binary addition being the exception. Out-of-place subroutines require a classical matrix as input. In-place subroutines involve additional classical preprocessing, specifically a PLU-decomposition, where the resulting matrices determine the circuit construction. This section is based on prior work, with some of the described subroutines detailed in~\cite{amento2013efficient, Banegas2020Concrete,kim2024toffoli}.

We use a polynomial basis representation, where $\mathbbm{F}_{2^n}$ is identified with $\mathbbm{F}_2[x]/p(x)$; where in this section we will use $p(x) \in \mathbbm{F}_2[x]$ to denote an irreducible polynomial of degree $n$. The elements in $\mathbbm{F}_{2^n}$ are then of the form
\begin{equation}
    f(x) = \sum_{i=0}^{n-1} f_{i}x^{i},
\end{equation}
where $f_{i} \in \mathbbm{F}_{2}$. Addition and multiplication are defined modulo an irreducible polynomial $p(x)$ of degree $n$. The irreducible polynomials $p(x)$ that are used in this work can be found in table~\ref{tab:irreducible_polynomials}. Using one qubit per coefficient of $ f(x) $, we encode $ f(x) $ as a $n$-qubit quantum state $\ket{f_{0}} \ket{f_{1}}\ldots\ket{f_{n-1}}$, which we collectively denote as $\ket{f}$; depending on the context, e.g., when referring to a subset of the $n$ qubits, we may make the sub-indices explicit, i.e., $\ket{f_{i,\ldots, j}}$ where $0\leq i<j\leq n-1$. 

\begin{table}[!ht]
\centering
\begin{tabular}{|l|l|}
\hline
$n$ & Irreducible polynomial           \\ \hline
$163$ & $x^{163} + x^7 + x^6 + x^3 + 1$  \rule{0pt}{10pt} \\ 
$233$ & $x^{233} + x^{74} + 1 $              \\ 
$283$ & $x^{283} + x^{12} + x^7 + x^5 + 1$ \\ 
$571$ & $x^{571} + x^{10} + x^5 + x^2 + 1$ \\ \hline
\end{tabular}
\caption{Irreducible polynomials of degree $n$ used in this work, taken from~\cite{kerry2013digital}.}
\label{tab:irreducible_polynomials}
\end{table}

\subsubsection{Out-of-place Multiplication}

In this section, we describe how to perform out-of-place multiplication. This operation involves multiplying $g(x) \in \mathbbm{F}_{2^n}$ by a fixed non-zero polynomial $h(x) \in \mathbbm{F}_{2^n}^{\ast}$, with the result reduced modulo an irreducible polynomial $p(x)$ of degree $n$. Since multiplication by a constant non-zero polynomial of $\mathbbm{F}_{2^n}$ is $\mathbbm{F}_{2}$-linear, this operation can be represented as a matrix-vector multiplication with a suitable $n\times n$ matrix $M$~\cite{amento2013efficient,kim2024toffoli}, where hereafter, the matrix and vector components are binary, and any operations over them are over $\mathbbm{F}_{2}$. Explicitly, $M$ encodes the multiplication by $h(x)\bmod{p(x)}$, where the $k$-th column of $M$ corresponds to the coefficients of $x^{k}\cdot h(x) \bmod{p(x)}$. The matrix $M$ acts on an $n$-dimensional column vector containing the coefficients of $g(x)$. This operation can be interpreted as adding coefficients of $g(x)$, conditioned on the elements of $M$.

To implement this in a quantum circuit, we start with the $n$-qubit input state $\ket{g}$, storing the coefficients of the polynomial $g(x)$, and an $n$-qubit state $\ket{f}$, initialized to store the coefficients of an arbitrary polynomial $f(x)$, where the result $g(x) h(x) \bmod{p(x)}$ will be output. Having expressed modular multiplication by a fixed polynomial $h(x)$ as a matrix-vector multiplication, we can realize this in a quantum circuit by applying CNOT gates, conditioned on the elements of the matrix $M$. Specifically, each $1$ in the matrix $M$ corresponds to applying a CNOT gate, where the column index specifies a control qubit of $\ket{g}$ and the row index specifies a target qubit of $\ket{f}$. The quantum circuit maps the input state $\ket{g}\ket{f}$ to the state $\ket{g} \ket{f + g\cdot h}$ and requires on average $n^2/2$ CNOT gates~\cite{kim2024toffoli}. However, in this work, we exactly count the number of CNOTs directly from $M$.

\subsubsection{Modular Reduction}

To compute the modular reduction of a $n-1$ degree polynomial $f(x)$ modulo a polynomial $p^{\prime}(x)$ of degree $d \leq n-1$, we note that we can express $f(x)\bmod{p^{\prime}(x)}$ as the sum of two polynomials 
\begin{equation}
     g(x)  +  \left( h(x)  \bmod{p^{\prime}(x)}\right),\label{Appendix:ModReduction2}
\end{equation}
where $g(x)$ is the polynomial consisting of terms of $f(x)$ of degree less than $d$, and the second polynomial represents the terms of $f(x)$ with degree greater than or equal to $d$, reduced modulo $p^{\prime}(x)$. \eqref{Appendix:ModReduction2} helps clarify how to compute the result using matrix-vector multiplication. Specifically, $ h(x) \bmod{p^{\prime}(x)} $ can be evaluated by applying a $ d \times (n-d) $ matrix $M$ to the column vector of coefficients of $ h(x) $, where the $k$-th column of $M$ is given by $ x^{k+d} \mod p^{\prime}(x) $. Consequently,~\eqref{Appendix:ModReduction2} can be interpreted as adding coefficients of $ h(x) $, conditioned on the elements of $M$, to the vector of containing coefficients of $ g(x) $.

To implement this in a quantum circuit, consider the $n$-qubit input state $\ket{f} = \ket{g}\ket{h}$, where $\ket{g},\ket{h}$ store the coefficients of the polynomials $g(x), h(x)$ as in (\ref{Appendix:ModReduction2}). Having expressed modular reduction as a matrix-vector multiplication, computing (\ref{Appendix:ModReduction2}) can similarly be implemented as in the previous subroutine. Explicitly, by applying CNOT gates conditioned on the elements of the matrix $M$,  with the control qubits in the $\ket{h}$ register and the target qubits in the $\ket{g}$ register. Similar ideas were first considered in~\cite{van2019space} and the case when the degree of $p'(x)$ can be smaller than $n-1$ is considered in~\cite{kim2024toffoli}. 

\subsubsection{In-place Addition}
In-place binary addition, which adds $ f(x) $ to another polynomial $ g(x) $, can be straightforwardly implemented as a quantum operation by applying $n$ CNOT gates to add $\ket{f}$ to an $ n $-qubit input state $\ket{g}$. In particular, this is a coefficient-wise XOR operation. The result of this addition, i.e., $\ket{f+g}$, replaces one of the input states, either $\ket{f}$ or $\ket{g}$, depending on the desired outcome.

\subsubsection{In-Place Multiplication}
In this section, we describe how to perform in-place multiplication, which refers to multiplication by a constant polynomial. This is the same setup as out-of-place multiplication, where we represent multiplying $g(x) \in \mathbbm{F}_{2^n}$ by a fixed non-zero polynomial $h(x) \in \mathbbm{F}_{2^n}^{\ast}$, modulo an irreducible polynomial $p(x)$ of degree $n$, by matrix-vector multiplication with a $n\times n$ matrix $M$. The $k$-th column of $M$ contains the coefficients of the polynomial $x^{k}\cdot h(x) \bmod{p(x)}$, with the coefficient of $x^0$ appearing in the first row.

Following~\cite{amento2013efficient,van2019space}, $M$ can be converted into a quantum circuit via a $PLU$-decomposition. This decomposition expresses $M$ as a product of three components: a permutation matrix $P$, a lower triangular matrix $L$, and an upper triangular matrix $U$, such that $M=PLU$. Such a decomposition allows for in-place multiplication, specifically:
\begin{itemize}
    \item The matrices $U$ and $L$ can be implemented as sequences of CNOT gates. Each off-diagonal element $1$ in $U$ and $L$ represents an application of a CNOT gate, where the column index indicates the control qubit and the row index indicates the target qubit.
    \item The permutation matrix $P$ can be implemented as a sequence of swap gates, where each off diagonal element $1$ in $P$, i.e., $P_{i,j} = 1$ (for $i\neq j$), represents a swap gate between qubits $i$ and $j$.
\end{itemize}

This in-place multiplication circuit requires at most $n^2 - n$ CNOT gates and a number of swaps~\cite{van2019space}. Note that in our resource estimation, we compute the exact CNOT and swap counts from $M$.

\subsubsection{In-Place Squaring}\label{appendix:sq}
In-place squaring, which involves squaring, modular reduction, and replacing the input, is a linear operation and can be expressed as a $n \times n$ matrix~\cite{amento2013efficient,Banegas2020Concrete}. Squaring $f(x)$ can be written as:
\begin{equation}
\left(\sum_{i=0}^{n-1} f_i x^i\right)^2=\sum_{i=0}^{n-1} f_i \cdot x^{2 \cdot i}.
\end{equation}
This operation can be expressed as a $(2n-1) \times n$ dimensional matrix acting on the coefficients of $f(x)$. To account for modular reduction, we can use a similar approach to that of the previous sections, and represent it as a $n\times (2n-1)$ matrix. Multiplying these two matrices yields a $n \times n$ that performs squaring and modular reduction. As in the in-place multiplication section, this $n\times n$ matrix can be efficiently converted into a quantum circuit using a $PLU$-decomposition. The decomposition expresses the matrix as a product of a permutation matrix $P$, a lower triangular matrix $L$, and an upper triangular matrix $U$, which can be implemented using swap gates and sequences of CNOT gates. This implementation has the same cost as the in-place multiplication.

When we need to perform $k$ consecutive squaring operations, we could naively perform each squaring operation separately; the cost scales linearly with $k$, requiring $k$ times the cost of squaring and modular reduction. An alternative approach involves generating the matrix for squaring and modular reduction once, and then multiplying this matrix by itself $k$ times, followed by a $PLU$-decomposition of the resulting matrix. The CNOT count of this approach does not grow with $k$. In practice, we will numerically evaluate both methods and choose the one with a smaller CNOT count.

\subsection{Modular Multiplication}\label{appendix:ModMultKKKH}

For modular multiplication, we use the CRT-based modular multiplication algorithm from~\cite{kim2024toffoli}, with some modifications. This algorithm is based on well-established classical techniques, combining Karatsuba-like recurrence formulas from~\cite{find2018better,ccalik2019searching} and the Chinese Remainder Theorem (CRT)~\cite{sunar2004generalized,fan2007comments}. Given two polynomials $f(x)$ and $ g(x)$  in $\mathbbm{F}_{2^n}$, the algorithm computes their product:
\begin{equation}
     f(x) g(x) \bmod{p(x)},
\end{equation}
where $p(x)$ is a degree-$n$ irreducible polynomial. It requires $n$ qubits to store each of $f(x)$ and $g(x)$, and $n$ qubits to store the result $h(x) + f(x) \cdot g(x)$, where $h(x) \in \mathbbm{F}_{2^n}$. In the following sections, we outline how the CRT can be used to perform modular multiplication. Next, we provide a step-by-step description of the algorithm, detailing how each step is implemented in a quantum circuit, with a necessary correction to the algorithm in the final step. Additionally, we describe the optimizations to the algorithm that were applied, the chosen input parameters, and their impact on the resource estimation.

\subsubsection{Modular Multiplication via CRT}
Let $f(x) = \sum_{i=0}^{n-1} f_i x^i $ and $g(x)=\sum_{i=0}^{n-1} g_i x^i$ be two binary polynomials of degree $n-1$. Suppose the aim is to compute their product
\begin{equation}
    r(x) = f(x)g(x) = \sum_{l=0}^{2n-2} c_l x^l, \label{eqn:classicalCRT_1}
\end{equation}
where $r_l=\sum_{i+j=l} f_i g_j$, for $0 \leq l \leq 2 n-2$. 
Directly computing this high-degree multiplication can be costly in terms of space and the number of AND operations~\footnote{In quantum circuits, the AND operation is implemented using a Toffoli gate.} required. To mitigate this, it is known that one can use a multiplication method based on the CRT for $\mathbbm{F}_2[x]$~\cite{sunar2004generalized,fan2007comments}:

\begin{theorem}\label{theorem:CRT}
\cite{fan2007comments} Let $m_1(x), m_2(x), \cdots, m_t(x)$ be pairwise co-prime polynomials and $m(x)=\prod_{i=1}^t m_i(x)$. Then, for any polynomials $r_1(x), r_2(x), \cdots, r_t(x)$, there is a unique polynomial $r(x) \bmod m(x)$, with $\operatorname{deg}(r(x))<\operatorname{deg}(m(x))$ such that 
\begin{equation}
r(x)=\sum_{i=1}^t r_i(x)q_i(x) \bmod m(x),\label{eqn:classicalCRT_2}
\end{equation}
where $r_i(x)=r(x) \bmod m_i(x)$ and 
\begin{equation}\label{eqn:crt_hi_inverse}
    q_i(x)=\left(\frac{m(x)}{m_i(x)}\right)\left(\left(\frac{m(x)}{m_i(x)}\right)^{-1} \bmod m_i(x)\right).
\end{equation}

\end{theorem}
Returning to the task of calculating the product in~\eqref{eqn:classicalCRT_1}: by Theorem~\ref{theorem:CRT}, to compute the product in~\eqref{eqn:classicalCRT_1}, we can equivalently compute the sum in~\eqref{eqn:classicalCRT_2}. Computing each term in the sum involves calculating the product $r_i(x)=f(x)g(x)  \bmod{m_i(x)}$, a multiplication by a constant polynomial $q_i(x)$, and modular reduction by $m(x)$. Furthermore, the product $r_i(x)=f(x)g(x)  \bmod{m_i(x)}$ can be further reduced to calculating $ f_{i}(x)g_{i}(x)  \bmod{m_i(x)}$, where $ f_{i} = f (x)  \bmod{m_i(x)}$ and $ g_{i} = g(x)  \bmod{m_i(x)}$. Therefore, we can see that we have reduced the problem of calculating the higher-degree polynomials to that of calculating the product of lower degree polynomials $f_{i}$  and $g_{i}$, both of degree less than $d_i = \text{deg}(m_{i}(x))$. Furthermore, for certain values of $d_i$, such as ($d_i \leq 8)$ efficient circuits that implement generalizations of the Karatsuba algorithm~\cite{find2018better,ccalik2019searching} can be applied to optimize the number of AND operations, implemented as Toffoli gates, required for modular multiplication.

\subsubsection{Quantum Circuit for CRT-Based Modular Multiplication}\label{app:qcircuit_crt}

\begin{figure}[!ht]
    \centering
    \begin{adjustbox}{width=1\textwidth}
    \begin{quantikz}[transparent]
    \input{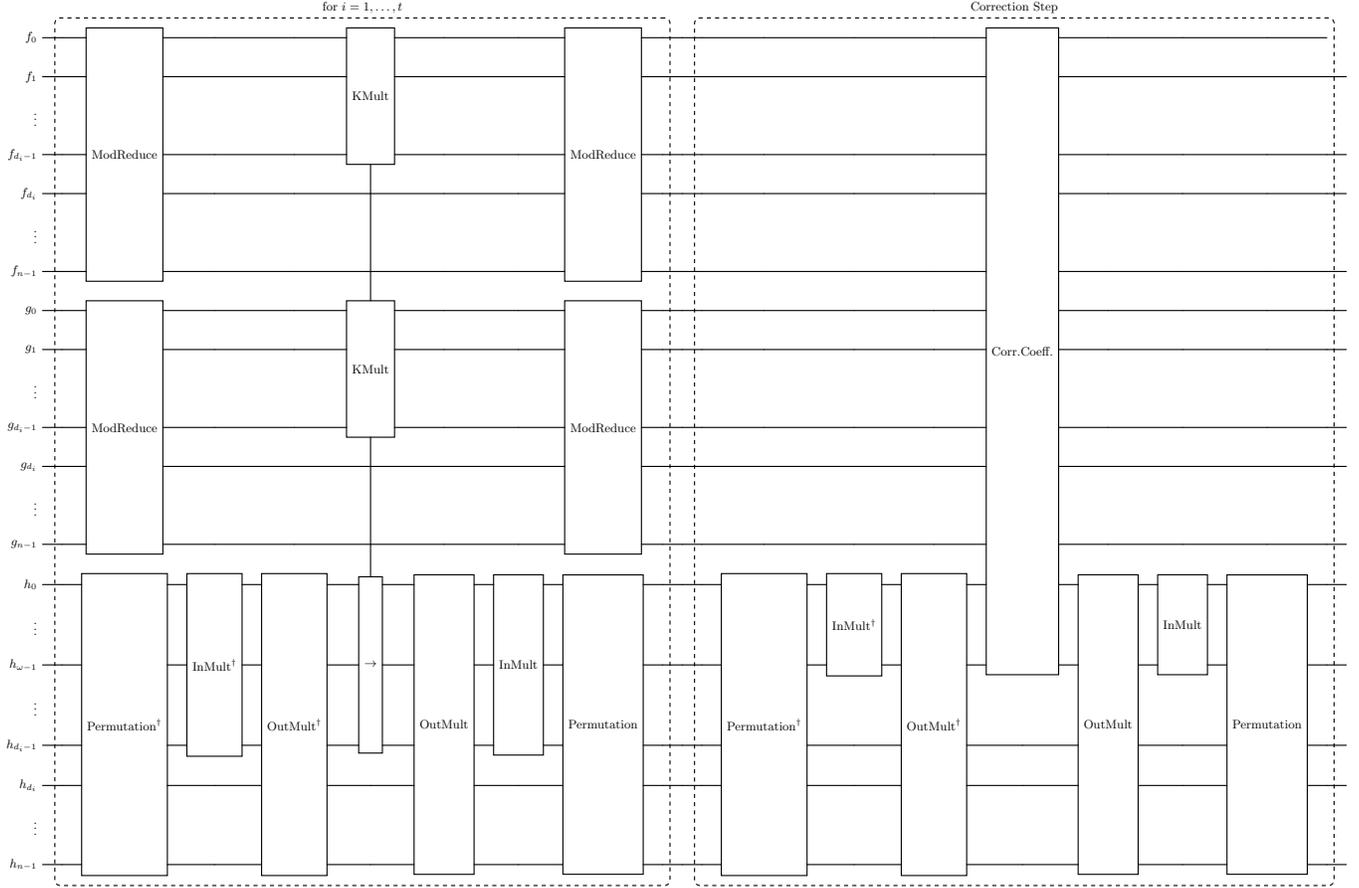}
    \end{quantikz}
    \end{adjustbox}
    \caption{Quantum circuit overview for the CRT-based modular multiplication algorithm~\cite{kim2024toffoli}. The circuit takes as fixed inputs the irreducible polynomial $p(x)$ of degree $n$ and the polynomials $m(x)=\prod_{i=1}^t m_i(x)$. It also takes as input two degree-$(n-1)$ polynomials $f(x)$ and $g(x)$, stored in separate $n$-qubit registers. Additionally, a target register initially stores the polynomial $h(x)$ and outputs $h(x) + f(x)g(x)\bmod{p(x)}$. The first dashed box corresponds to steps 1–3 of the algorithm for $i = 1,\ldots, t$, while the second dashed box performs step 4 for $\omega > 0$. ModReduce denotes the modular reduction subroutine, OutMult represents the out-of-place multiplication subroutine, and InMult corresponds to the in-place multiplication subroutine. These subroutines are described in section~\ref{appendix:toff_free}, with the required classical input for each explained in section~\ref{app:qcircuit_crt}. KMult represents modular multiplication, described in step 2, using generalized Karatsuba multiplication. However, if $deg(m_i(x))>8$ then this modular multiplication is replaced by a  recursive call to the CRT-based modular multiplication algorithm. Lastly, Corr.Coeff is the algorithm that computes the required correction coefficients, with the circuit shown in figure~\ref{fig:HighDegreeCircuit}. 
    }
    \label{fig:crt_modmult}
\end{figure}
Next, we explain the quantum circuit implementation of the CRT-based modular multiplication algorithm in four steps. Our implementation is largely based on~\cite{kim2024toffoli}; we will state explicitly where and why our implementation deviates. A high-level circuit diagram that implements the algorithm is provided in figure~\ref{fig:crt_modmult}.

The algorithm calculates the product 
\begin{equation}\label{eqn:crtmodmult-step-0}
    c(x) = f(x) g(x) \bmod{p(x)},
\end{equation}
where $p(x)$ is an irreducible polynomial of degree $n$. In the quantum circuit implementation, the inputs and outputs of the algorithm are as follows. The input to the quantum circuit consists of three binary polynomials $f(x), g(x), h(x) \in \mathbbm{F}_{2^n}$, stored in $n$-qubit states $\ket{f}$, $\ket{g}$ and $\ket{h}$, respectively, where  $\ket{h}$ serves as the target register for the result of the multiplication. The quantum circuit maps the input state $ \ket{f} \ket{g} \ket{h} $ to the output state $ \ket{f} \ket{g} \ket{h + f g \bmod{p} } $, where $p$ represents the irreducible polynomial $p(x)$ of degree $n$ provided as part of the algorithm's fixed inputs. In addition to the input polynomials, the algorithm requires a set of pairwise coprime polynomials $ m_1(x), m_2(x), \ldots, m_t(x)$, that define the polynomial $m(x)$ as their product:  
\begin{equation}\label{eqn:crt_defn_mx}
    m(x) = m_1(x) m_2(x) \ldots m_t(x).
\end{equation}
Denoting $d_i$ as the degree of each $m_i(x)$, and first computing the polynomials $ q_i(x)$ from each $m_i(x)$ as described in~\eqref{eqn:crt_hi_inverse}, the steps of the algorithm are then as follows:

\begin{description}
    \item[Step 1 Residue Computations] Calculate the residue representations of $f(x)$ and $g(x)$ with respect to each $m_i(x)$: 
    \begin{align}
    f^{i}(x) &=f(x) \bmod m_i(x), \label{eqn:ResidueComp1}\\
    g^{i}(x) &=g(x) \bmod m_i(x)\label{eqn:ResidueComp2}
    \end{align}
    for $i=1,\ldots, t$.

In the quantum circuit, these residue computations are performed using the modular reduction subroutine, described in section~\ref{appendix:toff_free}, with CNOT gates determined by the matrix $M_i$. Here, $M_i$ is a $ d_i \times (n-d_i) $ matrix constructed from the modulus polynomials $m_i(x)$, where each column $k$ contains the coefficients of $ x^{k+d_{i}} \bmod m_i(x) $, for $ k = 0, 1, \dots, n - 1 - d_i $. As shown in circuit implementation in figure~\ref{fig:crt_modmult}, for each $i$, the effects of this modular reduction must be uncomputed after Step 2. This is achieved by reapplying the same modular reduction subroutine with input $M_i$, using the fact that addition in $\mathbbm{F}_{2}$ is its own inverse. 
\end{description}

As done in~\cite{kim2024toffoli}, we optimize this step by adding the matrices $M_i$ (used for 
uncomputing the modular reduction for loop $i$) and $M_{i+1}$ (used for computing the modular reduction in the subsequent loop $i+1$). By appropriately padding these matrices with zeros (ensuring, in particular, that the column indices indicating the control qubits are correctly retained), they can be added into a single matrix $M_{i,i+1}$, which can used as input to the modular reduction subroutine and therefore reduces the CNOT count.

\begin{description}
\item[Step 2 Residue Products] Compute modular polynomial products
\begin{equation}
    c^{i}(x)=f^{i}(x) g^{i}(x)\left(\bmod m_{i}(x)\right)\label{eqn:ResidueProducts-1}
\end{equation}
for $i=1, \ldots, t$.
To compute the above equation in a quantum circuit there are two approaches. If the degree $d_i$  is greater than $8$, we recursively invoke the CRT-based modular multiplication algorithm.  However, if the degree $d_i$ of $m_i$, is less or equal to $8$, then we use Karatsuba-like formulae~\cite{montgomery2005five,weimerskirch2006generalizations}, i.e. generalized Karatsuba multiplication, to compute (\ref{eqn:ResidueProducts-1}). This method reduces the multiplication complexity (defined as the minimum number of multiplications needed to multiply two $n$-term polynomials) by recursively splitting a polynomial (in this case of degree $d_i$) into smaller parts and combining their products more efficiently. Standard Karatsuba multiplication uses $k=2$, while generalized Karatsuba multiplication applies $k$-way splits, further reducing the number of multiplications required. Moreover, it is possible to express generalized Karatsuba multiplication using a $v \times n$ matrix $T$ and a $l \times v$ matrix $R$, where $v$ is determined by the number of intermediate terms from the polynomial split and $l$  is the number of terms in the output. The matrices $T$ and $R$ are precomputed matrices, depending on the Karatsuba split that is used, and act on the $n$-dimensional column vectors containing the coefficients of $f(x)$ and $g(x)$~\cite{montgomery2005five,weimerskirch2006generalizations}:
\begin{equation}
    c = R\cdot[(T\cdot f)\circ(T\cdot g)],
\end{equation}
where $\cdot$ is the dot product and $\circ$ is the Hadamard product. In this work, we use $k$-way splits, where $k\in \left[3,8\right]$, and use the corresponding matrices $T$ and $R$, as these values have previously been optimized for minimal gate count in circuits~\cite{find2018better,ccalik2019searching}. Note that this is the step which incurs most of the Toffoli cost in the CRT-based modular multiplication algorithm. More specifically:
\begin{description}
\item[If $d_i \leq 8$] We invoke the generalized Karatsuba multiplication algorithm. This can be implemented via Algorithm 1 in~\cite{kim2024toffoli}, which maps the input state: 
\begin{equation}
    \ket{f^{i}}\ket{g^{i}}\ket{h}\mapsto\ket{f^{i}}\ket{g^{i}}\ket{h+ c^i},\label{eqn:karatsuba_map}
\end{equation}
where $c^i$ denotes the coefficients of the polynomial $c^i(x)$, of a smaller degree than $h(x)$, as defined in~\eqref{eqn:ResidueProducts-1}. The algorithm takes as input the precomputed matrices $T_{d_i}$, $R_{d_i}^{\prime}$, where $R_{d_i}^{\prime}$ is the modularly reduced matrix\footnote{That is, $R_{d_i}^{\prime}$ is obtained from a matrix that performs modular reduction, with respect to $m_i(x)$, acting on $R_i$.} of $R_{d_i}$. To determine the gate count, the algorithm proceeds as follows for each row of $T$. For each input register $\ket{f^{i}}$, $\ket{g^{i}}$, first find the minimum column index $i$ where the entry of $T$ is one. Then, for each column index greater than $i$, where the entry of $T$ is one, a CNOT gate is applied, with the control qubit at index $i$ and the target qubit at the column index. Next, for each column of $R$, identify the minimum row index $i$ where the entry of $R$ is one. For each row index greater than $i$, where the entry of $R$ is one, a CNOT gate is applied in the output register. Additionally, for each row in the matrix $T$, a Toffoli gate is then applied with controls qubits in $\ket{f^{i}},\ket{g^{i}}$ and  target qubits in $\ket{h}$. Finally, for every CNOT gate applied for the matrices $T$ and $R$, these must be uncomputed by running the sequence of CNOTs in reverse. This process is repeated for each row of $T$. We take the matrices $T_{d_i}$ and $R_{d_i}$ from the appendix of~\cite{kim2024toffoli}, and perform the modular reduction on $R_{d_i}$ by ourselves. Note that $T_{d_i}$ and $R_{d_i}$ stem from~\cite{montgomery2005five} for $d_i = 3$, ~\cite{find2018better} for $d_i \in \{4,5,6\}$ and~\cite{ccalik2019searching} for $d_i \in \{7,8\}$. In this work, we use $k\in \left[3,8\right]$ as these values have previously been optimized for minimal gate count in classical circuits~\cite{find2018better,ccalik2019searching}.
 
\item[If $ d_i > 8 $] To compute~\eqref{eqn:karatsuba_map}, we recursively call CRT-based modular multiplication algorithm with the following inputs: the polynomials $ f_i(x) $ and $ g_i(x) $ from Step 1 (each of degree less than $ d_i $), as well as the corresponding polynomial part of $ h(x) $. That is, the terms of $h(x)$ with degree less than $ d_i $. All operations are performed modulo $ m_i(x) $.  Lastly, the input includes a polynomial $ m^{\prime}(x) $, analogous to $ m(x) $ in Theorem~\ref{theorem:CRT}.
\end{description}
\end{description}
\begin{description}
\item[Step 3 CRT Polynomial] Compute the CRT-polynomial:
\begin{equation}\label{eqn:step3CRT_inversion}
c^{\prime}(x)=\sum_{i=1}^t \left(c^{i}(x) q_{i}(x)\bmod m(x)\right)\bmod p(x),
\end{equation}
where each quotient $q_i(x)$ can be computed through~\eqref{eqn:crt_hi_inverse}.

To evaluate each term, $\left(c^{i}(x) q_i(x)\bmod{m(x)}\right)\bmod p(x)$, in the above sum, this multiplication can be expressed as a matrix-vector multiplication using an $n\times d_i$ matrix $Q_{i}$, where each column $k$ contains the coefficients of the polynomial $ \left(x^{k} q_i(x)\bmod{m(x)}\right)\bmod{p(x)} $. Note that, the matrix $Q_{i}$ is a non-square matrix, so we cannot directly apply the in-place multiplication subroutine.

Nevertheless, this operation can be implemented in a quantum circuit as follows. First, perform a PLU decomposition on the matrix $Q_{i}$. This decomposition expresses  $Q_{i}$ as the product of three matrices: a $n\times n$ permutation matrix $P_{i}$, a  $n \times d_i$ matrix $L_{i}$, and a $d_i \times d_i$ matrix $U_{i}$, such that $Q_{i} = P_{i}L_{i}U_{i}$. By multiplying $L_{i}$ and $U_{i}$, we obtain a  $n \times d_i$ matrix $LU_{i}$. When then define two matrices: $M_i$ consisting of the first $d_i$ rows and $d_i$ columns of $LU_{i}$, and $N_i$, consisting of the last $n - d_i$ rows and $d_i$ columns of $LU_{i}$. Now to implement the operation 
\begin{equation}
Q_{i}=P_i\left[\begin{array}{c}
M_i  \\
N_i 
\end{array}\right],
\end{equation}
in a quantum circuit, we proceed by first implementing the $n - d_i \times d_i$ matrix $N_i$ via the out-of-place multiplication subroutine. The inputs are the control qubits $\ket{h_{0,\ldots,d_{i}-1}}$ and the targets qubits $\ket{h_{d_{i},\ldots,n-1}}$. Each off diagonal in $N_i$ corresponds to an application of a CNOT gate, where the column index where the column index indicates the control qubit and the row index indicates the target qubit. Next, $M_i$ can be realized by the in-place multiplication subroutine. The input is the $d_i$-qubit state $\ket{h_{0,\ldots,d_{i}-1}}$ and the $ d_i \times d_i$ matrix $M_i$ is implemented in the circuit via another PLU-decomposition. Lastly, the $n\times n$ permutation matrix $P_i$ can be implemented as a sequence of swap gates, where each off diagonal element $1$ in $P$, i.e., $P_{i,j} = 1$ (for $i\neq j$), represents a swap gate between qubits $i$ and $j$.

Additionally, in the quantum circuit for CRT-based modular multiplication, an inverse operation must be applied to the target register prior to step 2. This ensures that the multiplication $\ket{h + Q_{i}c^{i}}$ is performed, rather than inadvertently applying multiplication of $Q_{i}$ to any pre-existing values $h$ in the target register. More specifically, suppose that the vector $H$, corresponding to the coefficients of a polynomial, is stored in the $n$-qubit target register. To implement step 3 correctly, we must first apply the inverse of the circuit that implements the operation $Q_{i}$, followed by the circuit in step 2 that adds $K_i$, and finally apply the circuit for $Q_{i}$. It can be verified that this sequence ensures the desired result $H + Q_{i}K_i$, where $K_i$ is the vector with coefficients of (\ref{eqn:ResidueProducts-1}).

\end{description}

At this point, we have explained how steps 1-3 of the CRT-based multiplication algorithm can be vectorized and implemented in a quantum circuit. We further note that in the circuit implementation, we repeat steps 1-3 iteratively for $i= 1,\ldots, t$, resulting in the polynomial $c^{\prime}(x)$ being stored in the $n$-qubit target register. For the final step of the algorithm, we must consider two cases depending on the degree of the modulus polynomial $m(x)$, due to the challenges in selecting pairwise co-prime polynomials $m_i(x)$~\cite{sunar2004generalized}. These challenges arise from the need to ensure that the product
\begin{equation}
    m(x) = \prod_{i}m_i(x),
\end{equation}
has a sufficiently large degree relative to $n$, while also minimizing the degrees $d_i = deg(m_i(x))$ so that the generalized karatsuba algorithm is effectively employed. The simplest case to consider is when the modulus polynomial $m(x)$ is chosen so that $deg(m(x)) > 2n - 2$~\cite{sunar2004generalized}. In this case, the algorithm has computed the desired product in~\eqref{eqn:crtmodmult-step-0}, i.e.,
\begin{equation}
    c(x) = \left(c'(x) \bmod{m(x)} \right)\bmod{p(x)},
\end{equation}
and the algorithm terminates at this point. However, if $deg(m(x)) \leq 2n - 2$, then the computed product $c'(x)$ might not match the desired product $c(x)$. To see this, consider the case where $deg(m(x)) = 2n - 2$. When multiplying two polynomials $f(x)$, $g(x)$, each of degree $n-1$, their product can have a maximum degree of $2n-2$. Therefore, after step 3, the algorithm will produce the polynomial $c^{\prime}(x) = \left(f(x)g(x) \bmod{m(x)} \right)\bmod{p(x)}$ and would cause an undesired modular reduction of the term $x^{2n - 2}$. Fortunately, we can recover the desired product $c(x)$ by adding a term to $c^{\prime}(x)$\cite{sunar2004generalized}:
\begin{equation}
        c(x) = c'(x) + \left( c_{2n-2} m(x)\right) \bmod{p(x)},
\end{equation}
where $c_{2n-2} = f_{n-1}g_{n-1}$ is referred to as a correction coefficient. This correction process can be generalized for cases where $deg(m(x)) < 2n - 2$, which require multiple correction coefficients~\cite{fan2007comments}. The number of required correction coefficients is given by 
\begin{equation}
    \omega = 2n - 1 - deg(m(x)),
\end{equation}
and each correction coefficient can be calculated by the following formula: 
\begin{equation}\label{eqn:correction_coeffs-algebra}
    c_{2n - 2 - k} = 
    \sum_{i=n-1- k}^{n-1} s_{i} + 
    \sum_{\substack{i+j=2n-2-k,\\ n>i>j}} s_{i, j},
\end{equation}
where $k=0,\ldots,\omega - 1$, $s_i=f_i g_i$ and $s_{i,j}=\left(f_i+f_j\right)\left(g_i+g_j\right)$. For example, if the $deg(m(x)) = 2n - 3$ then two correction coefficients, $c_{{2n-2}}$ and $c_{2n-3}$, are required. Note that it is customary in the literature to indicate the use of a correction step by adding the symbolic term $(x-\infty)$ to $m(x)$.

\begin{figure}[!ht]
    \centering
    \includegraphics[width=\linewidth]{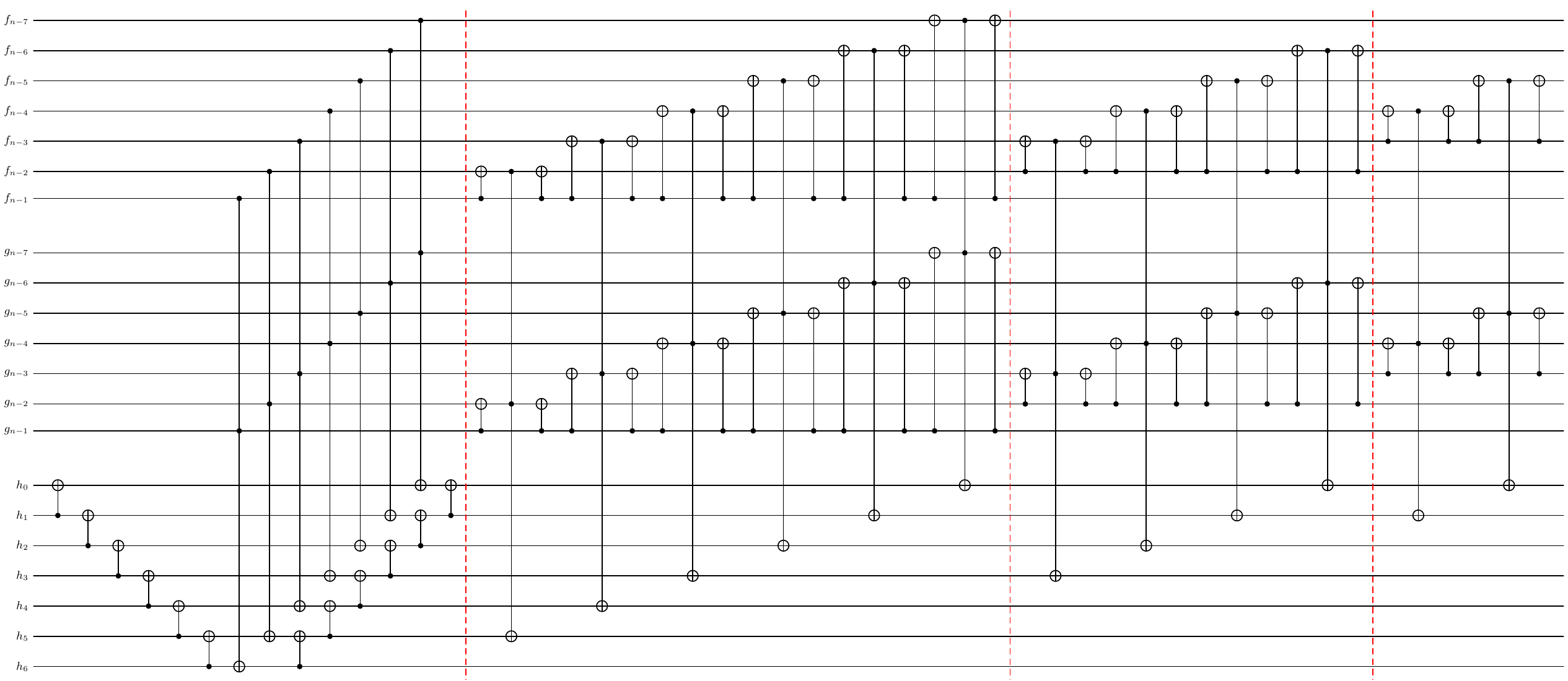}
    \caption{Quantum circuit for calculating correction coefficients for $\omega = 7$. The inputs to the algorithm are two polynomials $f$ and $g$ of maximum degree $n-1$, and a target register of size $\omega$ storing arbitrary coefficients $h_0, \ldots h_{\omega - 1}$. The output consists of the addition of correction coefficients $ c_{2n - 1 - \omega}, \ldots , c_{2n - 3}, c_{2n - 2}$ stored in the target register. The circuit works as follows: first, for all $k=0,\ldots,\omega - 1$, it computes the first sum in~\eqref{eqn:correction_coeffs-algebra}. After the first vertical dashed line in the circuit, the algorithm proceeds to compute the second sum in~\eqref{eqn:correction_coeffs-algebra}). Specifically, for each $k=0,\ldots,\omega - 1$, the algorithm starts by computing terms in the sum corresponding to $i = n-1$ satisfying $i+j = 2n - 2 -k$ (where $j = n-k-1$). After the next vertical line, where similarly for each $k=0,\ldots,\omega - 1$ and $i+j = 2n - 2 -k$, the circuit then computes terms for $i = n-2$ (where $j = n-k$), followed by $i = n-3$ (where $j = n-k+1$). This process is further repeated for subsequent indices $i$ and $j$ satisfying $i+j=2n-k-2$ and $n>i>j$. The final output of the circuit consists of the coefficients $h_0 + c_{2n-\omega - 1},$ $\ldots,$ $h_{\omega - 1} + c_{2n-2}$. }
    \label{fig:HighDegreeCircuit}
\end{figure}

We present a corrected quantum algorithm for calculating the required correction coefficients in~\eqref{eqn:correction_coeffs-algebra} for a specified 
$\omega$. Our algorithm for computing the correction coefficients is an in-place algorithm and works as follows: first, for all $k$, the algorithm computes the first sum in (\ref{eqn:correction_coeffs-algebra}), ensuring that the bitstring of the target qubits is preserved. Then, for all $k$, it proceeds to calculate the second sum, with details provided in the quantum circuit shown in figure~\ref{fig:HighDegreeCircuit}. This algorithm is based on the approach in~\cite{kim2024toffoli}. However, their circuit does not always preserve the existing bitstring in the target register. We have corrected their algorithm by adding the sequence of CNOTs preceding the first Toffoli gate and present an in-place circuit for calculating the correction coefficients. The algorithm requires $\omega + (\omega^2/4) $ Toffoli gates and $2(\omega - 1) + \omega^2$ CNOTs if $\omega$ is even, and $\omega + (\omega^2 -1 )/4 $ Toffoli gates and $2(\omega - 1) + \omega^2 - 1$ CNOTs if $\omega$ is odd. Additionally, we optimize our algorithm to reduce the CNOT count. Specifically, we minimize the CNOT gates required to calculate the second sum in~\eqref{eqn:correction_coeffs-algebra}: for each loop $i$ in the algorithm that computes the second sum, the CNOT gates immediately preceding the Toffoli gates can be propagated forward into a layer of CNOT gates, while the CNOT gates immediately following the Toffoli gates can similarly be propagated backward. Then by using the CNOT identity:
\begin{equation}
\begin{quantikz}
    & \targ{}   & \qw       & \targ{}   & \qw \\
    & \qw       & \targ{}   & \ctrl{-1} & \qw \\
    & \ctrl{-2} & \ctrl{-1}  & \qw      & \qw
\end{quantikz}
=
\begin{quantikz}
    & \targ{}       & \qw       & \qw   \\
    & \ctrl{-1}     & \targ{}   & \qw   \\
    & \qw           & \ctrl{-1} & \qw  
\end{quantikz}
\end{equation}
we can remove redundant CNOT gates between the calculations for successive indices $i$ and $i+1$. The optimized circuit requires $4(\omega - 1) + \omega^2/2$ CNOTs if $\omega$ is even, and $4(\omega - 1) + (\omega^2 - 1)/2$ CNOTs if $\omega$ is odd.

\begin{description}
    \item[Step 4 Correction] If $deg(m(x)) \leq 2n - 2$, then compute the remainder coefficients
    \begin{equation}\label{eqn:crtmodmult-step-4-1}
    c_{2n - 2},c_{2n - 3}, \ldots, c_{2n - 1 - \omega}
    \end{equation}
    for $\omega = 2n - 1 - deg(m(x))$ and compute the final product $c(x)$:
    \begin{equation}\label{eqn:crtmodmult-step-4-2}
    c(x)=c^{\prime}(x)+\sum_{i=2 n-1-w}^{2 n-2} c_i\left(\left(x^i\right)+\left(x^i \bmod m(x)\right)\right) \bmod p(x).
    \end{equation}
The implementation of the correction step in the quantum circuit is analogous to steps 2–3 of the algorithm. Specifically, the multiplication by the term $\left(\left(x^i\right)+\left(x^i \bmod m(x)\right)\right) \bmod p(x)$, in~\eqref{eqn:crtmodmult-step-4-2}, can be represented as a $n\times \omega$ matrix $H_\infty$. As in Steps 2–3, the quantum circuit that implements (\ref{eqn:crtmodmult-step-4-2}) is as follows. Run the circuit for $H_\infty$ in reverse to undo its action on the target register. Compute the correction coefficients in~\eqref{eqn:crtmodmult-step-4-1} using the circuit in figure~\ref{fig:HighDegreeCircuit}. Run the circuit that implements $H_\infty$.

The circuit that implements $H_\infty$ is similarly implemented via PLU-decomposition. As in step 3, $H_\infty$ is decomposed via a PLU decomposition and then equivalently rewritten in terms of a $n\times n$ permutation matrix $P_\infty$, a $w\times w$ matrix $M_\infty$ and a $(n-\omega)\times \omega$ matrix $N_\infty$. The matrix $H_\infty$ can then be realized in the circuit by an out-of-place multiplication (determined by $N_\infty$), an in-place multiplication subroutine (determined by $M_\infty$), and a sequence of swap gates determined by $P_\infty$.
\end{description}

We will now discuss the input choices for the quantum circuit and how they affect the resource estimates. The input to the CRT-based modular multiplication algorithm requires a modulus polynomial $m(x)$, which is the product of pairwise co-prime polynomials $m_i(x)$. How this $m(x)$ is chosen can affect the gate count and in particular the Toffoli count. Suppose the degree of $m(x)$ is chosen such that $deg(m(x))>2n-2$. If $n$ is large, e.g., $n = 283, 571$, attaining $deg(m(x))>2n-2$ may require introducing many higher-degree polynomials $m_i(x)$ with degrees exceeding 8. This could lead to multiple recursive calls to the algorithm, which would incur significant gate costs, particularly in terms of Toffoli gates. On the other hand, if $n$ is small, e.g., $n = 163, 233$, attaining $deg(m(x))>2n-2$ could be achieved by adding polynomials $m_i(x)$ with relatively small degrees. This approach would also incur gate costs resulting in calls to the generalized extended Karatsuba algorithm. For the $k(\in \left[3,8\right])$-way Karatsuba split used in the algorithm, the exact gate cost can be found in~\cite{kim2024toffoli}. Alternatively, suppose that the degree of $m(x)$ is chosen such that $deg(m(x))\leq 2n-2$. In this case, the extra cost comes from calls to the circuit for calculating correction coefficients, which becomes more expensive as the number of correction steps increases, leading to a higher count of Toffoli gates. Therefore, the choice of modulus polynomials impacts the overall Toffoli gate count; there is a trade-off between adding $m_i(x)$ and the cost of computing correction coefficients.

Repeated calls to the CRT-based modular multiplication algorithm, for large $n$, are costly compared to computing a few correction coefficients. For smaller $n$, one could increase the number of $m_i(x)$ polynomials to attain $deg(m(x))> 2n-2$, resulting in additional calls to the generalized Karatsuba algorithm. While correction steps can reduce costs, requiring more correction terms also adds to the gate cost. Thus, there is a trade-off in the Toffoli gate cost. A similar trade-off in active volume should apply as well. In practice, when choosing the polynomial $m(x)$, we have adopted a greedy and likely non-optimal approach, increasing the number of polynomials $m_i(x)$ until $deg(m(x))>2n-2$, whilst avoiding recursive calls to the algorithm in favour of correction steps where possible. This approach can be potentially be improved by choosing $m(x)$ more optimally in both Toffoli count and active volume, but we do not pursue it here as we anticipate the improvement to be minimal. Note that in the work of~\cite{kim2024toffoli}, for $n=283, 571$, the degree of $m(x)$ for the modulus polynomials listed in table~\ref{tab:mod_polys} do not satisfy the stated number of required correction coefficients. The polynomials $m(x)$ we have used in this work are listed in table~\ref{tab:mod_polys}. 

\begin{table}[!ht]
    \centering
    \begin{tabular}{c|l}
       $n$  &  Modulus Polynomials  \\\hline
        163  &  $ x^8,(x+1)^8, I_2^{1 \times 4}, I_3^{2 \times 2}, I_4^{3 \times 2}, I_5^{6 \times 1}, I_6^{9 \times 1}, I_7^{18 \times 1}, I_8^{7 \times 1}$  \\\hline 
         233  &  $ x^6,(x+1)^6, I_2^{1 \times 4}, I_3^{2 \times 2}, I_4^{33 \times 2}, I_5^{6 \times 1}, I_6^{9 \times 1}, I_7^{18 \times 1}, I_8^{25 \times 1}$  \\\hline 
          283  &  $x^7,(x+1)^6, I_2^{1 \times 4}, I_3^{2 \times 2}, I_4^{3 \times 2}, I_5^{6 \times 1}, I_6^{9 \times 1}, I_7^{18 \times 1}, I_8^{30 \times 1}, I_9^{6 \times 1}$ \\\hline 
           571  &  $x^9,(x+1)^8, I_2^{1 \times 4}, I_3^{2 \times 2}, I_4^{3 \times 2}, I_5^{6 \times 2}, I_6^{9 \times 1}, I_7^{18 \times 1}, I_8^{30 \times 1}, I_9^{56 \times 1}, I_{10}^{9 \times 1}$ 
    \end{tabular}
    \caption{Modulus polynomials used for CRT-based modular multiplication.  
    Let  $f_{d,i}(x)$ denote the \( i \)-th irreducible polynomial of degree \( d \), where \( i \) indexes the polynomials in arbitrary order. For integers \( a \) and \( b \), we define the set  $I_d^{a \times b} = \{ f_{d,i}(x)^b \mid i = 1,\ldots, a \}$.  The number of required correction coefficients is given by $\omega = 2n - 1 - deg(m(x))$. For $n = 283 $ and $n = 571$, we require $\omega = 4 $ and $\omega =  6$ correction coefficients, respectively. For $n = 163,233$, no correction step is required in the algorithm. Lists of irreducible polynomials are readily available online, e.g., see \url{https://mathworld.wolfram.com/IrreduciblePolynomial.html}. }
    \label{tab:mod_polys}
\end{table}

%%%%%%%%%%----------

\subsection{Modular Inversion}\label{appendix:InversionTT24}
The computation of modular inverses is the most resource-intensive arithmetic operation in elliptic curve point addition. In this work, we will compute inverses using algorithms based on Fermat's Little Theorem (FLT) as FLT-based inversion algorithms have significantly lower Toffoli gate costs compared to algorithms using the extended greatest common divisor (GCD)~\cite{Banegas2020Concrete,Taguchi2024On,Taguchi2023Concrete}. This method requires repeated use of the modular multiplication subroutine, which is costly in terms of Toffoli gates. However, this comes at the cost of needing more ancilla qubits, whereas GCD-based algorithms, while requiring fewer qubits overall, incur a much higher Toffoli gate count. In this work, we will use the inversion algorithm from~\cite{Taguchi2024On} because it has the lowest qubit count -- competitive to the GCD-based inversion algorithm -- among FLT-based inversion algorithms. The input to the algorithm is the state $\ket{f}\ket{0}^{R\cdot n}$ which is mapped to $\ket{f}\ket{f^{-1}}\ket{\text{garbage}}\ket{0}^n$, where $\ket{\text{garbage}}$ is of size $(R-2)n$ qubits and $R$ is a parameter dependent on the algorithm's classical input. The $n$-qubit states $\ket{f}, \ket{f^{-1}}$ encode the polynomials $f(x), f(x)^{-1}$, respectively. In the following sections, we outline how the FLT theorem can be used to compute modular inverses. Then, we describe the quantum implementation. Lastly, we discuss the optimizations applied to the algorithm, the chosen input parameters, and their impact on resource estimation.

\subsubsection{Modular Inversion via FLT}

Given a polynomial $f(x) \in \mathbbm{F}_{2^n}^{\ast}$, computing its inverse $f^{-1}(x) \in \mathbbm{F}_{2^n} $ can be computed via Fermat's Little Theorem~\cite{itoh1988fast} as:
\begin{equation}\label{eqn:FLT_theorem}
    f^{2^{n}-2} = f^{-1} \mod{p(x)},
\end{equation}
where $n$ is the degree of the irreducible $p(x)$ and we use the notation $f$ instead of $f(x)$ for convenience. Additionally, from here on, we will adopt the shorthand $f^x \coloneqq \langle x \rangle$.

Equation (\ref{eqn:FLT_theorem}) can be computed efficiently by the classical algorithm by Itoh and Tsujii~\cite{itoh1988fast}. This algorithm takes as input $f^1 =  \langle 2^{2^{0}}-1 \rangle$ and computes successive powers of $f$ through repeated squaring and multiplications. Specifically, it calculates $\langle 2^{2^{1}}-1 \rangle$, $\langle 2^{2^{2}}-1 \rangle$, \ldots $\langle 2^{2^{k_1}}-1 \rangle$, where $k_1=\lfloor\log (n-1)\rfloor$. Next, using the polynomials computed in this step, these intermediate results are combined to obtain $\langle 2^{n-1}-1 \rangle$, which is then squared to produce the desired polynomial $\langle 2^{n}-2 \rangle =  \langle -1 \rangle$. A quantum circuit that implements this method for computing inverses was first described in~\cite{amento2012quantum}, and requires $k_1 + t -1 $ Toffoli gates and $n \cdot \max \left(k_1+t-1, k_1+1\right)$ ancilla qubits., where $t\leq k_1 + 1$~\cite{Banegas2020Concrete}.

This method for computing inverses can be generalized to reduce the resource count — in some cases, reducing the Toffoli gate count~\cite{Taguchi2023Concrete} — or to reduce the ancilla qubits required~\cite{Taguchi2024On}. The key observation is that the exponents of 2, calculated in the Itoh and Tsujii algorithm, can be observed to be a specific addition chain\footnote{An addition chain for a non-negative integer $n-1$ is a sequence $\alpha_0 = 1, \alpha_1 , \alpha_2, \ldots, \alpha_l = n-1$, with the property that each $\alpha_i$, after $\alpha_0$, is obtained by adding two earlier terms (not necessarily distinct). The number $l$ is called the length of the addition chain.} for $n-1$. To see this, consider an example for $n = 163$. Using the Itoh and Tsujii algorithm, we would compute  $\langle 2^{2^{i}}-1 \rangle$ for $i = 1,\ldots, 7$, and then combine the previous polynomials to obtain $\langle 2^{2^{7}+2^{5}}-1 \rangle$ and 
$\langle 2^{2^{7}+2^{5}+2^{1}}-1 \rangle = \langle 2^{{n-1}}-1 \rangle$. By expressing the exponents of 2 as an addition chain for $n-1 = 162$, we have: 
\begin{align} 
&\left\{2^0,2^1, 2^2, 2^3, 2^4,2^5, 2^6,2^7, 2^7+2^5, 2^7+2^5+2^1\right\}\label{eqn:itoh_additionchain163}\\
= &\left\{1,    2,    4 ,   8 ,  16  , 32 ,  64  ,128 , 160,  162\right\} \nonumber.
\end{align} 
Note that the length $l$ of the addition chain (not including the first term) is $9$. There are seven terms $\left\{ 2, 4, 8, 16, 32, 64, 128\right\}$, which we refer to as doubled terms, and two terms,  $\left\{ 160,  162\right\}$ that we refer to as added terms. Every doubled term is computed by doubling its previous term, and every added term is computed by adding any two of its previous terms. The algorithms in~\cite{Taguchi2023Concrete,Taguchi2024On} allow the use of arbitrary addition chains to calculate $n-1$.  For example, consider an alternative addition chain for $n - 1 = 162$:
\begin{equation}\label{eqn:additionchain163}
    \{1, 2, 3, 6, 9,  18, 27, 54, 108, 162\}
\end{equation}
This chain also has a length of 9; however, note that it contains five doubled terms $\{2,6,18,54,108 \}$ and four added terms $\{3,9,27,162 \}$. In the next section, we will explain how addition chains are used to compute inverses in quantum circuits, which will clarify how properties of the addition chain affect the gate count. For now, it is important to note that the length of the chain $l$ corresponds to the number of modular multiplications required (i.e. Toffoli gates), and that computing doubled terms is more resource-intensive in terms of CNOT gates compared to computing added terms, which we discuss below. Thus, in this case, computing inverses with the addition chain in (\ref{eqn:additionchain163}) requires fewer CNOT gates compared to the addition chain  (\ref{eqn:itoh_additionchain163}).

However, a further observation can be made. Suppose each term $\alpha$ in the addition chain, corresponding to a polynomial $\langle 2^{\alpha}-1 \rangle$, must be stored in an $n$-qubit register within the circuit. Notice that some terms in the addition chain are not reused later in the computation. Consequently, these terms can potentially be cleared — i.e., the corresponding registers can be returned to the $\ket{0}^{n}$ state — and the cleared registers can then be reused to compute subsequent terms in the chain. For example, consider the addition chain for $n - 1 = 162$:
\begin{equation}\label{eqn:additionchain163_uncompute}
    \{1, 2, 3, 6, 9, 6, 3, 2, 18, 27, 54, 27, 18, 108, 162\}.
\end{equation}
In this addition chain, there are now decreasing terms that indicate
which polynomials can be cleared during the algorithm. According to (\ref{eqn:additionchain163_uncompute}), the algorithm takes as input the register that stores the polynomial $\langle 2^{1}-1 \rangle$ and then proceeds to compute the polynomials $\langle 2^{2}-1 \rangle,\langle 2^{3}-1 \rangle,\langle 2^{6}-1 \rangle, \langle 2^{9}-1 \rangle $, corresponding to $\{2,3,6,9\}$, which are stored in distinct $n$-qubit registers. Subsequently, the decreasing terms $\{6, 3, 2\}$ indicate that the registers storing  $\langle 2^{6}-1 \rangle,\langle 2^{3}-1 \rangle$ and $\langle 2^{2}-1 \rangle $ can be cleared. This is because the terms are not reused in the addition chain and, importantly, can be cleared using previously stored polynomials. Explicitly:
\begin{enumerate}
    \item $\langle 2^{6}-1 \rangle$, corresponding to $\{6 \}$ can be cleared using $\langle 2^{3}-1 \rangle$,  corresponding to $\{3 \}$.
    \item $\langle 2^{3}-1 \rangle$ can be cleared using the previously computed polynomials $\langle 2^{2}-1 \rangle$ and $\langle 2^{1}-1 \rangle$.
    \item $\langle 2^{2}-1 \rangle$ can be cleared using the polynomial $\langle 2^{1}-1 \rangle$.
\end{enumerate}
Similarly, the terms  $\{ 27, 18 \}$ indicate that registers storing $\langle 2^{27}-1 \rangle,\langle 2^{18}-1 \rangle$ can also be cleared. However, this clearing process—i.e., returning a register to the zero state for reuse—comes at a cost. Each term that is cleared requires a modular multiplication and may also involve additional squaring operations. However, the upshot of this clearing process is significant: for a slight increase in the number of multiplication operations, we can approximately halve the number of ancilla qubits required for inversion for our values of $n=163,233,283,571$. The total number of required ancilla qubits is $R n$, where $R = (2l - \Tilde{l} + 1)$,  $\Tilde{l}$  is the length of the  addition chain (excluding the initial term 1), and $l$ is the number of strictly increasing terms in the addition chain (also excluding the initial term 1). Note that, the clearing process can be omitted to further reduce the Toffoli count~\cite{Taguchi2023Concrete}, but this significantly increases the number of ancilla qubits required. The resource estimates for this inversion algorithm without the clearing process are given in table~\ref{tab:TT23inversion_results}.

\begin{table}[!ht]
    \centering
    \begin{tabular}{c|l}
       $n$  &  Addition Chains \\\hline
        163  &  $\{1, 2, 3, 6, 9, 6, 3, 2, 18, 27, 54, 27, 18, 108, 162\}$ \\
         233  &  $\{1, 2, 3, 4, 7, 4, 3, 2, 14, 28, 29, 28, 14, 58, 116, 58, 232\}$ \\
          283  &  $\{1, 2, 3, 6, 9, 15, 9, 6, 3, 30, 45, 47, 45, 30, 2, 94, 141, 94, 282\}$\\
           571  &  $\{1, 2, 3, 4, 7, 4, 3, 2, 14, 28, 29, 57, 29, 28, 14, 114, 171, 285, 171, 114, 570\}$  
    \end{tabular}
\caption{Addition chains for values of $n$ used in this work for the inversion algorithm from~\cite{Taguchi2024On}. The chains were taken directly from~\cite{Taguchi2024On} and include terms that indicate which register can be cleared in the quantum circuit.  } 
    \label{tab:addition_chains}
\end{table}

\subsubsection{Quantum Circuit for FLT-Based Modular Inversion with Addition Chains}

\begin{figure}[t]
    \centering
    \begin{adjustbox}{width=\textwidth}
    % \begin{quantikz}
    % \input{quantikz_circuits/inversion_circuit}
    \includegraphics[width=\linewidth]{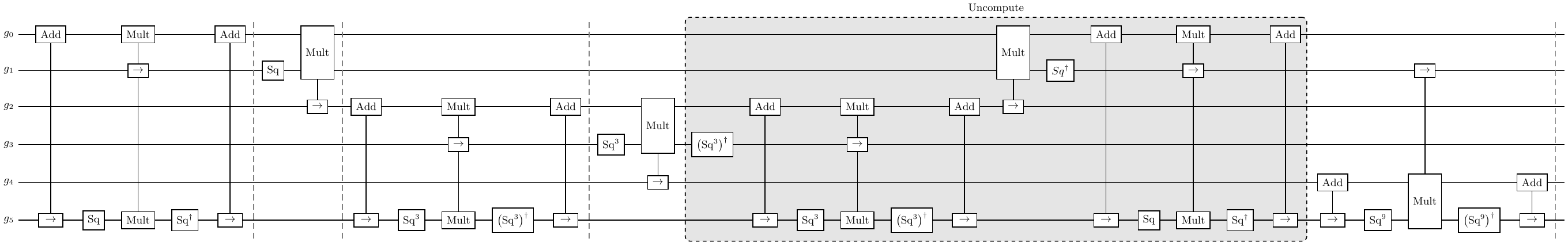}
    % \end{quantikz}
    \end{adjustbox}
    \begin{adjustbox}{width=\textwidth}
    % \begin{quantikz}
    % \input{quantikz_circuits/part2inversion}
    \includegraphics[width=\linewidth]{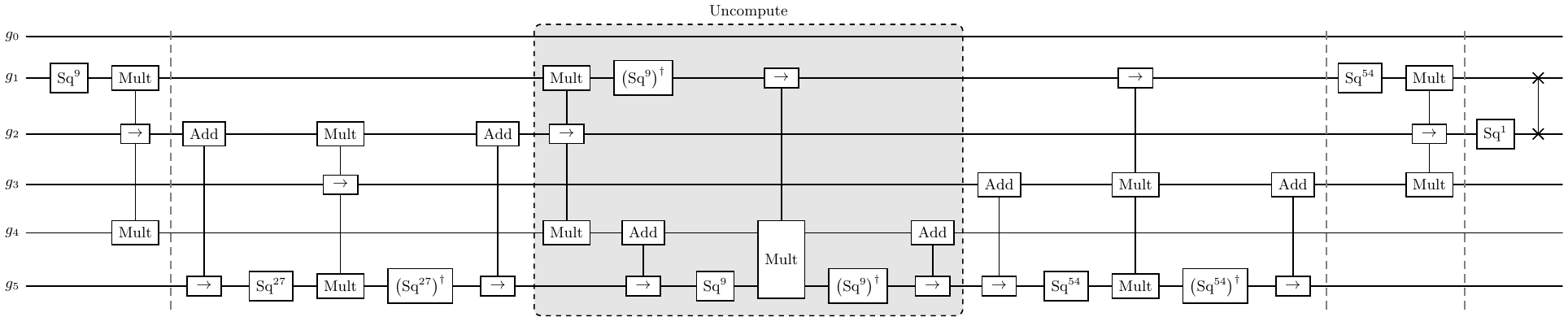}
    % \end{quantikz}
    \end{adjustbox}
    \caption{Quantum circuit for computing the modular inverse for $n=163$ using the FLT-based inversion algorithm from~\cite{Taguchi2024On}. The circuit is split into two parts: the top circuit represents the first half, while the bottom circuit corresponds to the second half. The circuit takes as input an irreducible polynomial $p(x)$ of degree $n$ and a polynomial $f(x) \in \mathbbm{F}_{2^n}^{\ast}$. The input polynomial $f(x)$ is stored in register $g_0$, and its computed inverse $f(x)^{-1}$ is stored in the $g_1$ register. The algorithm requires $R$ $n$-qubit ancilla registers, where $R = 2l - \Tilde{l} + 1$. At the end of the algorithm, the $g_5$ register is cleared to the $\ket{0}^n$ state.  The addition chain used in this example can be found in table~\ref{tab:addition_chains}. As in figure~\ref{fig:s123}, the ``$\rightarrow$" marks the target registers whose values are modified by modular multiplication and addition. The squaring operation is denoted by ``Sq", and $k$ consecutive squaring operations are denoted ``Sq$^k$". Each vertical dashed line indicates the computation of a doubled or added term in the addition chain. The highlighted sections of the circuit correspond to the clearing steps of the algorithm, specifically the terms $\left\{ 6, 3, 2 \right\}$ and $\left\{ 27, 18 \right\}$ in the given addition chain example.
    }
    \label{fig:tt24_163}
\end{figure}

In this section, we outline how to implement the quantum circuit for FLT-based modular inversion with addition chains~\cite{Taguchi2024On}. We provide an explicit inversion circuit for $n=163$ in figure~\ref{fig:tt24_163}.

To compute $f^{-1}$, the inversion algorithm~\cite{Taguchi2024On} takes as input an addition chain of $\Tilde{l}$, excluding the first term, and the state $\ket{f}\ket{0}^{R\cdot n}$. The output state is  $\ket{f}\ket{f^{-1}}\ket{\text{garbage}}\ket{0}$, where $\ket{\text{garbage}}$ is of size $(R-2)n$. The number of ancilla qubits required is given by $R n$, where $R = (2l - \Tilde{l} + 1)$. The number of modular multiplications is determined by $\Tilde{l}$, and the number of CNOT gates depends on the added and doubled terms in the addition chain. At the end of the inversion algorithm, the first ancilla register contains the result  $f^{-1}$, and an ancilla register is cleared to the $\ket{0}^n$ state. This cleared register will be exploited and use to store intermediate arithmetic results (see section~\ref{sec:algo}).

To explain the algorithm, and without loss of generality, consider the following addition chain:
\begin{equation}\label{eq:rb_addition_chain}
\left\{\alpha, \beta ,\gamma, \delta, \gamma,\beta \right\}, 
\end{equation}
where $\alpha, \beta ,\gamma, \delta$ are non-negative integers such that $\alpha < \beta < \gamma < \delta$, and $\beta$ is a doubled term ($\beta = 2\alpha$), $\gamma$ an added term such that $\gamma = \alpha + \beta $, and $\delta$ a doubled term ($\delta = 2\gamma$). We now demonstrate how each term $\nu$ in the addition chain — whether a doubled or added term — corresponding to $\langle 2^{\nu}-1 \rangle$, is computed and then cleared within a quantum circuit~\cite{Taguchi2023Concrete,Taguchi2024On}.%
\begin{description}
    \item[Computing a doubled term] Suppose we have a term $\alpha$ in the addition chain that we want to double, i.e., $\beta = 2\alpha$. This corresponds to a register $g_i$ (for some index $i$) storing the polynomial $\langle 2^{\alpha}-1 \rangle$. To compute $\beta = 2\alpha$, proceed as follows. Using the binary addition subroutine, add the polynomial stored in register $g_i$ to $h$, where $h$ is initially in the $\ket{0}^n$ state, resulting in $h = \langle 2^{\alpha}-1 \rangle$. Next, apply the squaring operation to $h$, $\alpha$ number of times, resulting in $h = \langle 2^{\alpha}-1 \rangle^{2^\alpha}$. Then, apply modular multiplication to the polynomials stored in $g_i$, $h$, outputting the result in a register $g_{i+1}$, which is in the state $\ket{0}^n$. This results in: 
       \begin{equation}
           g_{i+1} = \langle 2^{\alpha}-1\rangle^{2^\alpha} \langle 2^{\alpha}-1 \rangle = \langle 2^{2\alpha}-1\rangle,
       \end{equation}
    where $\beta = 2\alpha$ is desired doubled term in the addition chain. Lastly, to clear $h$, apply the circuit for squaring in reverse, $\alpha$ times, and add $g_i$ to $h$. This last step ensures that $h$ is returned to its initial state $\ket{0}^n$ and can be reused.
    %
    %added terms!
    \item[Computing an added term] Suppose we terms $\alpha$ and $\beta$ in the addition chain that we want to add, where this corresponds to a register $g_i$ storing $ \langle 2^{\alpha}-1\rangle$ and $g_{i+1}$ storing $\langle 2^{\beta}-1\rangle$, respectively. To compute $\gamma = \alpha + \beta$, proceed as follows. Apply the squaring operation to $g_{i+1}$, $\alpha$ number of times, resulting in $g_{i+1} = \langle 2^{\beta}-1 \rangle^{2^\alpha}$. Then, apply modular multiplication to the polynomials stored in $g_i$ and $g_{i+1}$, outputting the result in a register $g_{j}$, (which initially is in the state $\ket{0}^n$). This results in: 
        \begin{equation}
            g_{j} = \langle 2^{\alpha}-1 \rangle \langle 2^{\beta}-1\rangle^{2^\alpha}  = \langle 2^{\alpha + \beta}-1\rangle,
        \end{equation}
    where $\gamma = \alpha + \beta$ is the desired added term in the addition chain. Notice that the register $g_{i+1}$ now stores $\langle 2^{\beta}-1\rangle^{2^\alpha}$, therefore, if $\beta$ is reused, we would need to undo the squaring operations.
\end{description}
The addition chain in (\ref{eq:rb_addition_chain}) indicates that we can clear, in order, the registers containing $\left\langle 2^\gamma-1\right\rangle$ and $\left\langle 2^\beta-1\right\rangle$, corresponding to the terms $\gamma$ and $\beta$. The clearing process, returning a register to the $|0\rangle^n$ state -- entails reversing the operation used to compute the term. Specifically, this involves, potentially squaring (or an inverse squaring) operation, reapplying the corresponding operation (whether doubling or addition) and then performing binary addition. Since adding a polynomial to itself twice results in zero, the register is effectively reset to the $|0\rangle^n$ state. This process can be performed as long as the terms in the addition chain are not reused later, and the corresponding polynomial remains present and stored in the quantum circuit at the time of clearing.

\begin{description}
    \item[Clearing an added term] Suppose we want to clear the register  $g_{j}$ containing the polynomial corresponding to the added term $\gamma = \alpha+\beta$. We have that $\langle 2^{\beta}-1 \rangle^{2^\alpha}$ is stored in register $g_{i+1}$, and $\langle 2^{\alpha}-1 \rangle$ is stored in register $g_i$, from the previous steps of the algorithm\footnote{If this is not the case, e.g.. for a different addition chain where the register $g_{i+1}$ contains, $\langle 2^{\beta}-1 \rangle^{2^k}$ for some integer $k$, then apply the necessary squaring (or inverse) operation to ensure the exponent is $2^{\alpha}$, i.e $\langle 2^{\beta}-1 \rangle^{2^\alpha}$.}. To clear the register $g_{j}$, apply modular multiplication to the polynomials stored in $g_{i}$, $g_{i+1}$, outputting the result in the register $g_{j}$, which contains the polynomial $\langle 2^{\gamma}-1 \rangle$. This results in: 
    \begin{equation}
    g_{j} =  \langle 2^{\gamma}-1\rangle + \langle 2^{\gamma}-1\rangle = 0,
    \end{equation}
    hence, the register $g_j$ is cleared to the $\ket{0}^n$ state.
    \item[Clearing a doubled term] Suppose we want to clear the register containing the polynomial corresponding to the doubled term $\beta = 2\alpha $. From the previous steps of the algorithm, $\langle 2^{\beta}-1 \rangle^{2^\alpha}$ is stored in register $g_{i+1}$, and $\langle 2^{\alpha}-1 \rangle$ is stored in register $g_i$. 
    To clear the register, $g_{i+1}$ we first apply the circuit for squaring in reverse, $\alpha$ times, resulting in $g_{i+1}=\langle 2^{\beta}-1 \rangle$. Then the process is similar to that of computing a doubled term. Add the polynomial stored in register $g_i$ to $h$, where $h$ is initially in the $\ket{0}^n$ state. Next, apply the squaring operation to $h$, $\alpha$ number of times. Then, apply modular multiplication to the polynomials stored in $g_i$ and $h$, outputting the result in the register $g_{i+1}$: \begin{equation}
    g_{i+1} =  \langle 2^{\beta}-1\rangle + \langle 2^{\beta}-1\rangle = 0.
    \end{equation}
    This has now cleared the $ g_{i+1}$ register to the $\ket{0}^n$ state.
\end{description}

The above description shows how to compute/uncompute each term (either a doubled or added term) in the addition chain in a quantum circuit. The last term in the addition chain corresponds to calculating $\langle 2^{n-1}-1\rangle$. Lastly, to compute the inverse, we simply square the term $\langle 2^{n-1}-1\rangle$ to obtain $\langle 2^{n}-2\rangle \equiv \langle -1\rangle$.

We briefly comment on optimizations performed in the resource estimation. In the algorithm, there are many repeated consecutive calls to the modular squaring operation—for example, to double the term  $\gamma$, one would need to perform $\gamma$ squaring operations (as well as running the inverse of this circuit to clear the register). As discussed in section~\ref{appendix:toff_free}, there are two approaches one could use: perform each squaring operation separately in the circuit or combine the consecutive squaring operations into one operation, and then implement this in the circuit. In our resource estimation, we numerically calculated both methods and chose the method with a smaller CNOT count. A similar optimization was also applied in~\cite{Taguchi2024On}, though it is unclear to us whether our method coincides with theirs. The resource estimates for the FLT-based inversion algorithm from~\cite{Taguchi2024On}, using the addition chains in table~\ref{tab:addition_chains} and the clearing process, are summarized in table~\ref{tab:inversion_results}. Note that the gate counts reported in~\cite{Taguchi2024On} are derived using the circuits from~\cite{kim2024toffoli} without our corrections and optimizations in the modular multiplication routine.

Lastly, we note that the Toffoli count in the inversion algorithm can be reduced if the clearing process is omitted~\cite{Taguchi2023Concrete}. In this case, one can use the algorithm from~\cite{Taguchi2023Concrete}, which takes as input $\ket{f}\ket{0}^{l\cdot n}$ and outputs $\ket{f}\ket{\text{garbage}}\ket{f^{-1}}$, with $\ket{\text{garbage}}$ is of size $(l-1)n$. This approach reduces the number of modular multiplications to $l$ but increases the ancilla qubit requirement to $l\cdot n$. The resource estimates for this inversion algorithm, using the addition chains in table~\ref{tab:addition_chains} without the clearing process, are provided in table~\ref{tab:TT23inversion_results}. Note that the gate counts reported in~\cite{Taguchi2023Concrete} are derived using the circuits from~\cite{kim2024toffoli} without our corrections and optimizations in the modular multiplication routine.

\begin{table}[!ht]
\begin{tabular}{|l|l|l|l|l|l|}
\hline
$n$ &  $\#$ ModMults  & $\#$ Toffolis & $\#$ CNOTs & $\#$ Swaps & Active Volume \\ \hline
163 & 9        & 8991   & 1096546 & 13265   & $4.81 \times 10^6$      \\ \hline
233 & 10       & 14480  & 2408816    & 52610   & $1.03 \times 10^7$      \\ \hline
283 & 11       & 19536  & 3977687   & 43671  & $1.68 \times 10^7$    \\ \hline
571 & 12       & 46320  & 16058155  & 114550  & $6.64 \times 10^7$     \\ \hline
\end{tabular}
\caption{The costs of computing $ f^{-1}(x)\bmod{p(x)}$ given $f(x)$ via the FLT-based inversion algorithm~\cite{Taguchi2023Concrete}, using the addition chains in table~\ref{tab:addition_chains} without the clearing process. The costs are stated in terms of the number of modular multiplication applications, Toffolis, CNOTs, and swaps, as well as active volume. In this approach, the ancilla qubit requirement is $l\cdot n$, where $l$ is the number of strictly increasing terms in the addition chain (excluding the initial term 1) and is equal to the number of modular multiplications.}
\label{tab:TT23inversion_results}
\end{table}

\end{document}